\documentclass[fleqn,usenatbib]{mnras}

\usepackage[T1]{fontenc}

\DeclareRobustCommand{\VAN}[3]{#2}
\let\VANthebibliography\thebibliography
\def\thebibliography{\DeclareRobustCommand{\VAN}[3]{##3}\VANthebibliography}

\usepackage{graphicx}	
\usepackage{amsmath}	
\usepackage{amssymb}	

\title[Observed structural properties of EAGLE galaxies]{Observed structural parameters of EAGLE galaxies: reconciling the mass-size relation in simulations with local observations}

\author[A. de Graaff et al.]{
Anna de Graaff$^{1}$\thanks{E-mail: graaff@strw.leidenuniv.nl},
James Trayford$^{2}$,
Marijn Franx$^{1}$,
Matthieu Schaller$^{3,1}$,
Joop Schaye$^1$,
Arjen van der Wel$^4$
\\
$^{1}$Leiden Observatory, Leiden University, P.O.Box 9513, NL-2300 AA Leiden, The Netherlands\\
$^{2}$Institute of Cosmology \& Gravitation, University of Portsmouth, Dennis Sciama Building, Burnaby Road, Portsmouth PO1 3FX, UK\\
$^{3}$Lorentz Institute for Theoretical Physics, Leiden University, PO Box 9506, NL-2300 RA Leiden, The Netherlands\\
$^{4}$Sterrenkundig Observatorium, Universiteit Gent, Krijgslaan 281 S9, B-9000 Gent, Belgium
}

\date{Accepted XXX. Received YYY; in original form ZZZ}

\pubyear{2021}

\usepackage{newtxtext,newtxmath}

\begin{document}
\label{firstpage}
\pagerange{\pageref{firstpage}--\pageref{lastpage}}
\maketitle

\begin{abstract}

We use mock images of $z=0.1$ galaxies in the 100\,Mpc EAGLE simulation to establish the differences between the sizes and morphologies inferred from the stellar mass distributions and the optical light distributions. The optical, $r$-band images used were constructed with a radiative transfer method to account for the effects of dust, and we measure galaxy structural parameters by fitting S\'ersic models to the images with \textsc{Galfit}. 
We find that the derived $r$-band half-light radii differ systematically from the stellar half-mass radii, as the $r$-band sizes are typically $0.1$\,dex larger, and can deviate by as much as $\approx0.5\,$dex. The magnitude of this size discrepancy depends strongly on the dust attenuation and star formation activity within the galaxy, as well as the measurement method used. Consequently, we demonstrate that the $r$-band sizes significantly improve the agreement between the simulated and observed stellar mass-size relation: star-forming and quiescent galaxies in EAGLE are typically only slightly larger than observed in the GAMA survey (by 0.1\,dex), and the slope and scatter of the local mass-size relation are reproduced well for both populations. Finally, we also compare the obtained morphologies with measurements from GAMA, finding that too few EAGLE galaxies have light profiles that are similar to local early-type galaxies (S\'ersic indices of $n\sim 4$). Despite the presence of a significant population of triaxial systems among the simulated galaxies, the surface brightness and stellar mass density profiles tend to be closer to exponential discs ($n\sim1-2$). Our results highlight the need to measure the sizes and morphologies of simulated galaxies using common observational methods in order to perform a meaningful comparison with observations.

\end{abstract}

\begin{keywords}
galaxies: evolution -- galaxies: structure -- galaxies: stellar content -- galaxies: fundamental parameters
\end{keywords}



\section{Introduction}\label{sec:intro}

The sizes and morphologies of galaxies are some of their most basic observable properties, and provide crucial insight into the formation of galaxies and the build-up of their stellar mass. Cosmological hydrodynamical simulations that aim to model a realistic universe are therefore expected to reproduce such fundamental characteristics. However, to determine the success of a given model requires a fair comparison between simulations and observations, as the latter can come with significant biases due to the systematic differences between the distribution of the light and the stellar mass.

Observationally, galaxy morphologies are highly diverse, but are usually grouped into two classes, of early-type (spheroidal or bulge-like) and late-type (more disc-like) systems. Importantly, these morphological types have been found to correlate with other properties: early-type galaxies are typically more massive than late-type galaxies, have significantly redder colours and lower star formation rates \citep[e.g.,][]{Blanton2003,Kauffmann2003,Driver2006}, and are often found to lie in denser environments \citep[e.g.,][]{Dressler1980,Gomez2003}. Early-type galaxies thus appear to have followed very different evolutionary paths from late-type galaxies, although the precise mechanisms behind the quenching of star formation in galaxies and the possible link to a morphological transformation still represents an active area of research.

Furthermore, the stellar masses and sizes of both populations of galaxies have been shown to be correlated at low redshift \citep[e.g.,][]{Shen2003,Lange2015}, and this relation has been observed to exist at least up to $z\sim3$ \citep[e.g.,][]{Trujillo2006,vdWel2014a,Mowla2019}. The sizes of late-type galaxies can be linked back to the dependence of the halo angular momentum on halo mass \citep{Mo1998}. To zeroth order, the galaxy size reflects the size of the halo, but it further depends on the details of more complex processes, such as stellar feedback \citep[e.g.,][]{Sales2010,Brook2011,DeFelippis2017}, or the formation of a central bulge component through mergers or gravitational instabilities \citep[e.g.,][]{Hernquist1989,Dekel2014,Zolotov2015}. For early-type galaxies, the mass-size relation is much steeper and evolves faster than is the case for the late-type population, suggesting a different formation history. Dry mergers are thought to play a significant role \citep{Naab2009,Bezanson2009}, and \citet{Shen2003} demonstrated that a simple model in which galaxies undergo repeated minor mergers, can describe both the slope and scatter of the observed mass-size relation of quiescent galaxies at $z\sim 0$ well.

As the stellar mass-size relation reflects fundamental processes in the formation and evolution of galaxies, it provides a key measure of success for theoretical models, and cosmological hydrodynamical simulations in particular. The latest generation of cosmological simulations all approximately reproduce the observed mass-size relations, e.g., the EAGLE simulations \citep{Schaye2015,Furlong2017}, Illustris-TNG \citep{Genel2018}, or SIMBA \citep{Dave2019}. Moreover, these simulations are able to form a diverse set of morphologies, as both star-forming discs and quiescent spheroids are formed \citep[e.g.,][]{Snyder2015,Correa2017,Thob2019}.

However, many of these studies are based on a comparison between the stellar mass distributions of simulated galaxies, and the optical light observed in photometric galaxy surveys. Additionally, there are often differences in the measurement techniques used: galaxy sizes in simulations are typically measured using a curve of growth method, whereas observational studies tend to fit parametric models to estimate galaxy sizes.

To mitigate possible biases introduced in these comparison studies, much effort has gone into the post processing of simulations to produce realistic mock observations. At the core, these mock data all consist of optical images, which are created by modelling the spectral energy distributions (SEDs) of the stellar particles to estimate the total light emitted within a specified wavelength range. Further possible layers of complexity are the addition of a sky background and photon noise, and modelling of the effects of dust. Even without the inclusion of dust attenuation, these mock images have demonstrated the importance of colour gradients: sizes measured from simulated, optical images are generally larger than the corresponding stellar mass sizes \citep{vdSande2019}, which is in line with observational findings \citep[e.g.,][]{Szomoru2013,Mosleh2017,Suess2019}. The mass-size relation is therefore also changed, and simulated galaxies are found to be larger than observed \citep{Snyder2015,Bottrell2017b,vdSande2019}, although the galaxy populations in Illustris-TNG show relatively good correspondence with observations \citep{Genel2018,Lin2021}.

Most of the aforementioned studies, however, do not measure galaxy size in the same manner as observational studies, or do not model the effects of dust in their mock images. More progress on the latter front has been made in studies that measure galaxy morphologies from mock images created with radiative transfer codes, which model the dust absorption and scattering of light between the point of emission and an observer \citep[e.g., \textsc{sunrise}, \textsc{skirt}, or \textsc{powderday}][]{Jonsson2006,Baes2011,Camps2015,Narayanan2021}. With these more realistic images, \citet{Rodriguez2019} (Illustris-TNG) and \citet{Bignone2020} (EAGLE) found that galaxy morphologies at $z\sim0$, as quantified by non-parametric methods \citep[for a review, see][]{Conselice2014}, agree well between simulations and observations.

To also make the measurement of galaxy sizes consistent with observations, requires fitting the mock surface brightness profiles with S\'ersic models \citep{Sersic1968}. These models are highly instructive, as they simultaneously measure the overall scale (size, luminosity) and morphology of a galaxy (quantified by the S\'ersic index and the projected axis ratio). On the other hand, the modelling of S\'ersic profiles is strongly dependent on the estimation and treatment of the sky background and noise within an image, and dedicated software for the robust extraction of structural parameters has therefore been developed \citep[e.g., \textsc{Galfit}, \textsc{gim2d};][]{Peng2002,Simard2002}. 
Using such software, \citet{Price2017} demonstrated the importance of the measurement method used on the inferred size, as the sizes of high-redshift galaxies in the MassiveFIRE simulations differ significantly between measurements with \textsc{Galfit} and aperture-based methods. With a custom fitting method, \citet{Rodriguez2019} found that the $z\sim 0$ mass-size relation in Illustris-TNG depends only weakly on the method used to measure the half-light radius, but the S\'ersic profile sizes of the simulated galaxies appear to be systematically larger than equivalent measurements from the Pan-STARRS $3\pi$ Steradian Survey.

Clearly, there are many factors at play when comparing simulations and observations: the physics implemented in the simulation (and the limited fidelity thereof), the level of `realism' of the forward modelled mock data, and consistency in the analysis methods used. In this work, we aim to perform a consistent comparison between the structural properties of galaxies in the EAGLE simulation and galaxies from the Galaxy And Mass Assembly survey \citep[GAMA;][]{Driver2011,Liske2015,Baldry2018}. We measure the structural parameters of the simulated galaxies using near-identical methods to large galaxy surveys, and do so for the projected stellar mass distributions, as well as optical images that include dust attenuation \citep[from][]{Trayford2017}. This allows us to not only perform a robust comparison with observations from GAMA, but also to distinguish between the effects of colour gradients and differences in the measurement methods used. 

We first describe the EAGLE simulations and the construction of the optical images used in Section~\ref{sec:data}. Section~\ref{sec:methods} discusses the subsequent creation of realistic mock images that include instrumental effects and noise, as well as the S\'ersic profile modelling and associated quality control. We compare different measures of galaxy size in Section~\ref{sec:results_sizes}, and demonstrate how both the adopted measurement method and colour gradients (due to stellar population gradients and dust) within galaxies affect the overall mass-size relation. The morphological properties obtained with the S\'ersic profile modelling are presented in Section~\ref{sec:results_morph} and compared with observations from GAMA.  Finally, we discuss the implications of our findings in Section~\ref{sec:discussion}, and summarise our key results in Section~\ref{sec:conclusion}.

\section{Data}\label{sec:data}

\subsection{EAGLE simulations}\label{sec:eagle_description}
The \textsc{EAGLE} simulations consist of a suite of smoothed particle hydrodynamics (SPH) simulations for a range of different volumes, resolutions, and subgrid models \citep{Schaye2015,Crain2015}. Here, we use the reference model run for the largest available comoving volume of $100^3\,\rm Mpc^3$ (L100N1504), which assumes a flat $\Lambda$CDM cosmology with cosmological parameters obtained from \citet{Planck2014}: $\Omega_{\rm m}=0.307$, $\Omega_{\rm b}=0.0482$ and $H_0 = 67.77\,\rm km\,s^{-1}\,Mpc^{-1}$. This simulation has a mass resolution of $9.7\times10^6\,{\rm M_\odot}$ for the dark matter particles, and $1.81\times10^6\,{\rm M_\odot}$ for the initial mass of the gas particles. As a result, galaxies of stellar mass $M_*\gtrsim 10^{10}\,{\rm M_\odot}$ are typically resolved by $\gtrsim 10^4$ stellar particles at $z\sim0$. The Plummer-equivalent gravitational softening scale is $\epsilon=0.70$\,proper kpc at $z<2.8$, and the gravitational force starts to get softened on scales smaller than $2.8\epsilon\approx2\,$kpc. From hereon, we will use proper lengths for all quoted distances and sizes, unless stated otherwise.

Haloes are identified in EAGLE using the friends-of-friends algorithm, and self-bound substructures within haloes are identified using the \textsc{subfind} algorithm \citep{Springel2001,Dolag2009}. 
We follow the convention of \citet{Schaye2015} and define galaxies as the collection of particles that belong to a single substructure, with the galaxy stellar mass defined as the sum of the stellar particles enclosed within a spherical aperture with radius $30\,$kpc centred on the potential minimum.
We focus our analysis on galaxies at $z=0.1$ (snapshot 27), for which mock optical imaging created with \textsc{skirt} is available, as described in Section~\ref{sec:skirt}. Given the limited spatial resolution in the simulation, we impose a lower limit on the stellar mass of $M_* = 10^{10}\,{\rm M_\odot}$\,, as \citet{Ludlow2019} showed that galaxies below this stellar mass tend to have sizes smaller than the convergence radius of the dark matter, leading to the spurious transfer of energy from dark matter to stars via 2-body scattering.
Selecting all galaxies of stellar mass $M_* \geq 10^{10}\,{\rm M_\odot}$ from the public EAGLE database \citep{McAlpine2016}, we obtain a sample of 3624 galaxies.

\subsection{Galaxy images}\label{sec:skirt}

Optical images, presented in \citet{Trayford2017}, were generated by post-processing the EAGLE data with \textsc{skirt} \citep{Baes2003,Baes2011,Camps2015}. 
The principle of the radiative transfer code \textsc{skirt} is to trace monochromatic `photon packages' from a source to a specified detector using a Monte Carlo method. In this way, unlike with the commonly adopted method of applying a dust screen, representative 3D absorption and scattering of light due to dust are accounted for, thus creating a realistic image. 
We provide a brief summary of these data below, and refer the reader to \citet{Camps2016} and \citet{Trayford2017} for a detailed description of the procedures involved.

The stellar particles in the snapshot, provided they lie within a 30\,kpc radius around the centre of the galaxy, form the source of the photon packages. As described in \citet{Trayford2015}, each particle older than $>100\,$Myr is treated as a single stellar population, and assigned a SED with \textsc{GALAXEV} \citep{Bruzual2003}, using the initial mass, metallicity and stellar age from the simulation snapshot and assuming a \citet{Chabrier2003} initial mass function. The spatial distribution of the light emitted by the particle is described by a truncated Gaussian distribution, with a smoothing length dependent on the distance to the 64th nearest neighbour. 

For younger stars, the additional absorption by dust in the birth clouds needs to be taken into account. Given the limited mass resolution, however, this firstly requires a resampling of the recent star formation of the stellar particles with young ages ($<100$\,Myr), which is done in a similar fashion to \citet{Trayford2015}. Sub-particles older than $10\,$Myr are treated as described above, whereas younger populations are instead assigned SEDs using the \textsc{MAPPINGS-III} code \citep{Groves2008}, which models the emission and dust absorption within HII regions. The smoothing length for these young populations is taken to be dependent on their mass and the local gas density, although the net kernel (i.e., including the position) is equivalent to that of the other stellar particles. We note that the choice of the smoothing lengths sets the level of granularity in the final images, and \citet{Bignone2020} showed that this likely affects some of the non-parametric morphological measurements. However, in Appendix~\ref{sec:smoothing_lengths} we demonstrate that the smoothing has a negligible effect on the parametric morphologies measured in this work.

Dust in the diffuse interstellar medium (ISM) is modelled based on the properties and spatial distribution of the gas particles in the galaxy. Gas particles are smoothed using the SPH smoothing lengths, and the ISM is then discretised over an adaptive grid with a minimum grid cell size of 0.11\,kpc. The dust mass within the grid cells is calculated from the star-forming as well as the cold ($T<8000\,$K) gas mass, by assuming a constant dust-to-metal mass ratio \citep{Camps2016}. Dust mass in the HII regions, already implemented through the \textsc{MAPPINGS-III} SEDs, is also accounted for. The composition of the dust grains is taken as the model by \citet{Zubko2004}, a multi-component interstellar dust model that provides a good fit to the observed extinction curve of the Milky Way, as well as the diffuse infrared emission and abundance constraints.

With the source of emission and distribution of the dust defined, the \textsc{skirt} calculations are performed on a finely sampled wavelength grid (333 wavelengths in the range $0.28-2.5\micron$), resulting in an integral field data cube. Broadband imaging is constructed by convolving the cube with an instrument response function and integrating along the wavelength direction. This is done for both the observed ($z=0.1$) and rest frame for three different projections: face-on, edge-on, and random (the projection along the $z$-axis of the simulation box). 
Images have a field of view of $60\times 60\,$kpc$^2$ with a pixel scale of $0.234\,\rm kpc\,pix^{-1}$, which at $z=0.1$ corresponds to an angular resolution of $0\farcs123\,$pix$^{-1}$.

\section{Methods}\label{sec:methods}

\subsection{SDSS mock images} \label{sec:mock_sdss}

The images generated with \textsc{skirt} provide a realistic view of the optical emission of the simulated galaxies. However, unlike real observations of galaxies, these images do not include any instrumental effects or background noise. We therefore use the randomly orientated \textsc{skirt} images as a starting point to construct mock observations, specifically, to mimic typical image data from the Sloan Digital Sky Survey (SDSS). We choose to focus on only the $r$-band images \citep[in the observed frame;][]{Doi2010}, as this is the wavelength range commonly used in observational studies.

In addition to creating mock SDSS images of the optical light, we construct mock `images' of the stellar mass distributions directly from the simulation snapshot. These stellar mass maps are designed to have similar noise properties and resolution as the optical imaging, to allow for a robust comparison between the distributions of the optical emission and stellar mass.

\subsubsection{Optical images}\label{sec:mock_ugriz}

The initial images are $60\,$kpc on a side, which in most cases is significantly larger than the half-mass radius of the galaxy. However, more massive galaxies can have large half-mass radii ($>10\,$kpc), or contain extended star-forming discs. Although a $30\,$kpc aperture may capture all the galaxy mass, at least for systems of $M_* < 10^{11}\,{\rm M_\odot}$ \citep{Schaye2015}, a large spatial extent can still be problematic in the S\'ersic modelling, as the extended emission may get mistaken for background flux.
We therefore add empty background pixels onto the sides of the images, such that they become $60\arcsec\times60\arcsec$ in area ($114^2\,\rm kpc^2$). 

Next, we add a uniform background and convolve the image with a Gaussian point-spread function (PSF) to match the sky background and seeing in the SDSS imaging. We calculate the median value of the $r$-band `sky' and `psfWidth' from the photometric field catalogue of the ninth data release of the SDSS \citep[DR9;][]{sdss:dr9}, to set the sky background level ($\mu_{\rm sky} = 20.9\,$mag\,arcsec$^{-2}$) and the full width at half maximum ($\rm FWHM=1\farcs39$) of the PSF, respectively. 
We note that the real PSF in the SDSS image data has a far more complex shape than the single Gaussian profile assumed here. However, in Appendix~\ref{sec:apdx_psf} we show that a simple PSF model is sufficient for measuring parametric morphologies.
After the convolution, we resample the image from a pixel scale of $0\farcs123\,\rm pix^{-1}$ to $0\farcs396\,\rm pix^{-1}$ to match the SDSS pixel resolution.

From the same SDSS DR9 catalogue we obtain typical values for the $r$-band detector gain ($G=4.73\,e^-\,\rm ADU^{-1}$), the conversion factor from counts to fluxes (nMgyPerCount\,$=0.0051\,\rm nmgy\,ADU^{-1}$), and the `dark variance' (the combination of detector readout noise and the dark current; $\sigma^2_{\rm dark}=1.32\,\rm ADU^2$). The dark variance is added to the image to mimic detector effects, under the assumption that these electrons follow a Poisson distribution (i.e., $\mu=\sigma^2$), although this source of noise is insignificant in comparison to the sky background level (a factor $\approx20$ lower). Lastly, we convert the image to units of $e^-\,\rm pix^{-1}$.

\begin{figure}
    \centering
    \includegraphics[width=\linewidth]{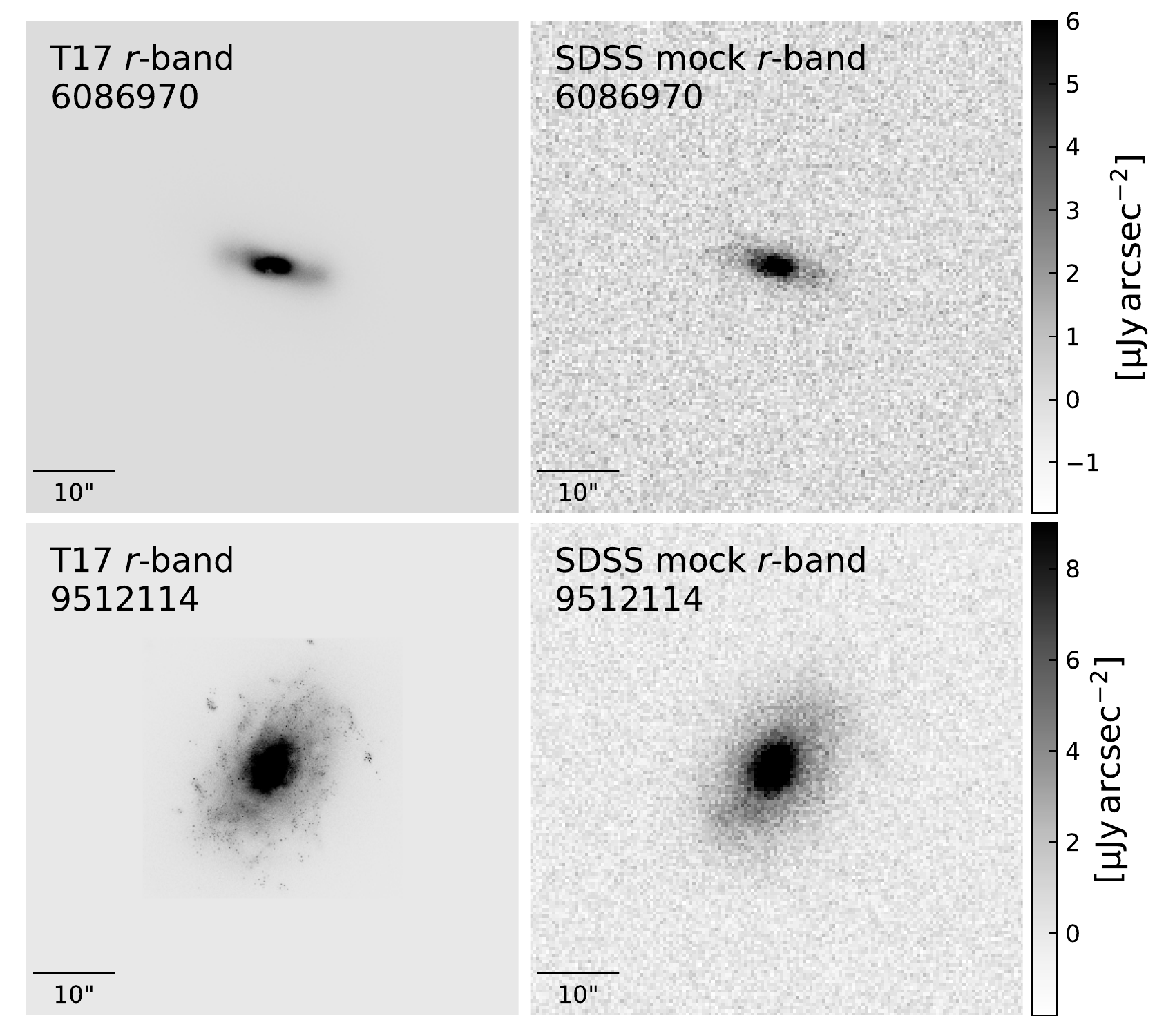}
    \caption{Examples of the $r$-band images constructed with \textsc{skirt} of galaxies at $z=0.1$ \citep[{left};][]{Trayford2017}, and the corresponding mock SDSS images that include realistic instrumental and sky effects ({right}). }
    \label{fig:example_skirt}
\end{figure}

The image now closely resembles the collection of photoelectrons by a detector, and these photoelectrons obey Poisson statistics. We can therefore create an image with a realistic noise level: for each pixel, we draw a random sample from the Poisson distribution with mean value equal to the number of electrons in that pixel ($\mu=N_{\rm e, pix}$). We also obtain a `sigma image', an image with the same dimensions as the galaxy image that stores $G^{-1}\times\sqrt{N_{\rm e, pix}}$, which will be used as statistical weights in the two-dimensional S\'ersic modelling (Section~\ref{sec:sersic}). We note that the image construction with \textsc{skirt} (Section~\ref{sec:skirt}) also introduces Poisson noise, however, this noise is well below the typical noise level in the SDSS \citep{Trayford2017}, therefore justifying the seemingly duplicate addition of photon noise.

As a final step, we divide the image by the gain and subtract the (previously added) sky background and dark variance from the noisy image, delivering the final mock SDSS image. Fig.~\ref{fig:example_skirt} shows an example of an initial $r$-band image created with \textsc{skirt}, and the corresponding mock SDSS image (converted to physical flux units) that includes realistic noise and PSF smoothing.

Unlike the real SDSS data, these mock images do not contain any foreground or background sources, as only light from within a $30\,$kpc aperture is included. We have chosen to not implement this additional complexity, as \citet{Bottrell2017a} showed that the effect of crowding on the measurement of structural parameters is generally small, with the exception of very low surface brightness systems that are few in number.

\subsubsection{Stellar mass images}\label{sec:mock_im_mstar}

To construct images of the stellar mass distribution that match the noise and image resolution properties of the $r$-band images, we follow a similar methodology to the previous section, with few modifications. Rather than starting from the \textsc{skirt} data, we begin from the EAGLE particle data and select a box of size $114^3\,\rm kpc^3$ centred around the potential minimum of the galaxy. Within this box, we select only the stellar particles that are identified as being part of the galaxy by the \textsc{subfind} algorithm. In this way, analogous to the \textsc{skirt} images, neighbour galaxies are not included in the images. The current stellar mass of these particles is then projected in the $x-y$ plane of the simulation box to obtain an image of $512\times512$ pixels, which is the same orientation and of similar spatial resolution as the \textsc{skirt} data.

To be able to add realistic noise as described in Section~\ref{sec:mock_ugriz}, an effective mass-to-light ratio ($\Upsilon_{\rm eff}$) is required that describes the typical scaling between the $r$-band and stellar mass imaging. To obtain $\Upsilon_{\rm eff}$, we first compute the ratio ($\Upsilon$) between the stellar mass of the galaxy (within the spherical aperture of radius $30\,$kpc), and the observed flux within a circular aperture of $30\,$kpc in the noise-free optical images. We then use the median of this distribution, $\Upsilon_{\rm eff} = 10^{11}\,{\rm M_\odot}\rm\,mJy^{-1} $\,, to convert the stellar mass images to an effective flux and hence to a number of photoelectrons. We use a fixed value of $\Upsilon_{\rm eff}$ for all galaxies, as the variation in $\Upsilon$ is relatively small: the standard deviation of $0.13\,$dex in $\Upsilon$ corresponds to variations in the image noise level of $\sim 15\%$, which we have found the S\'ersic profile fitting procedure (Section~\ref{sec:galfit}) to not be sensitive to. 

\begin{figure}
    \centering
    \includegraphics[width=\linewidth]{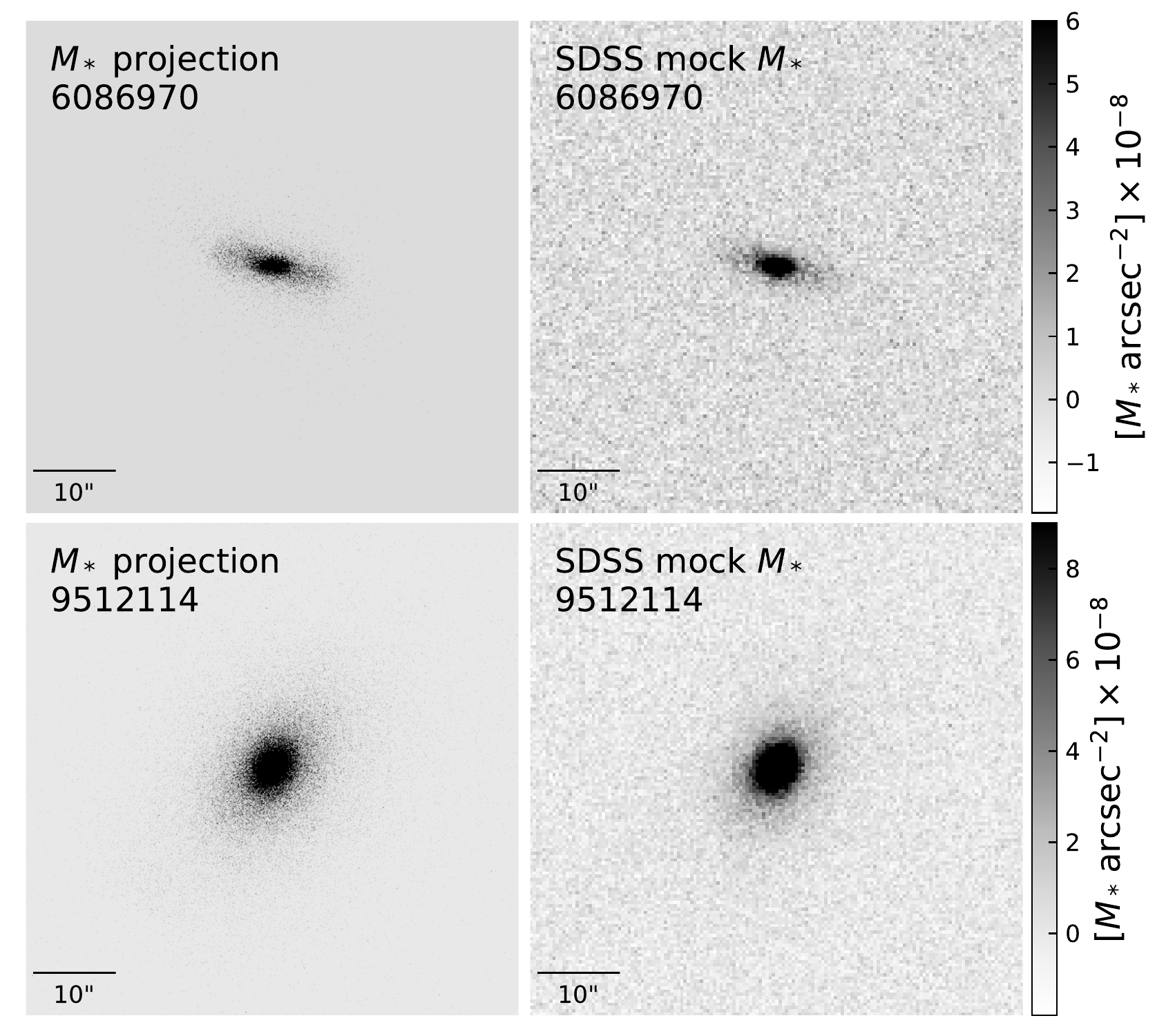}
    \caption{Examples of the stellar mass maps of galaxies at $z=0.1$ created by projecting the stellar particles along the $z$-axis of the simulation box ({left}), and the corresponding mock SDSS `stellar mass images' that include realistic instrumental and sky effects ({right}).}
    \label{fig:example_massmap}
\end{figure}

As in Section~\ref{sec:mock_ugriz}, we add a uniform sky background level and smooth the image with a Gaussian PSF, using the same $\mu_{\rm sky}$ and PSF FWHM as before. The image is then resampled to a pixel scale of $0\farcs396\,\rm pix^{-1}$, and the dark variance is added. We apply a Poisson noise model, and subtract the total (sky + dark variance) background to produce our final mock SDSS image of the stellar mass distribution. Fig.~\ref{fig:example_massmap} shows an example of the initial $x-y$ projection of the stellar particles, and the mock SDSS image.

\subsection{S\'ersic modelling}\label{sec:sersic}

We model the light and stellar mass profiles of the simulated galaxies by fitting a two-dimensional, parametric model to the mock imaging. This model, a single S\'ersic profile \citep{Sersic1968}, is described by five parameters: the total AB magnitude ($m$) or stellar mass ($M_{\rm *,S{\acute e}rsic}$), the S\'ersic index ($n$), the half-light or half-mass semi-major axis ($r_{\rm e,maj}$), the ratio of the semi-major and semi-minor axes ($q$), and the position angle ($\phi$).

We describe our fitting procedure in detail in the following sections. In summary, we use a combination of \textsc{SExtractor} \citep{Bertin1996} and \textsc{Galfit} \citep{Peng2010} to estimate the initial values of the S\'ersic parameters and to find the best fitting parameter values, respectively. Both softwares are commonly used in observational studies that measure structural parameters of galaxies \citep[e.g.,][]{Barden2012,Kelvin2012,vdWel2012,Meert2015}, which enables us to perform a consistent comparison between simulated and observational results.

\begin{figure*}
    \centering
    \includegraphics[width=\linewidth]{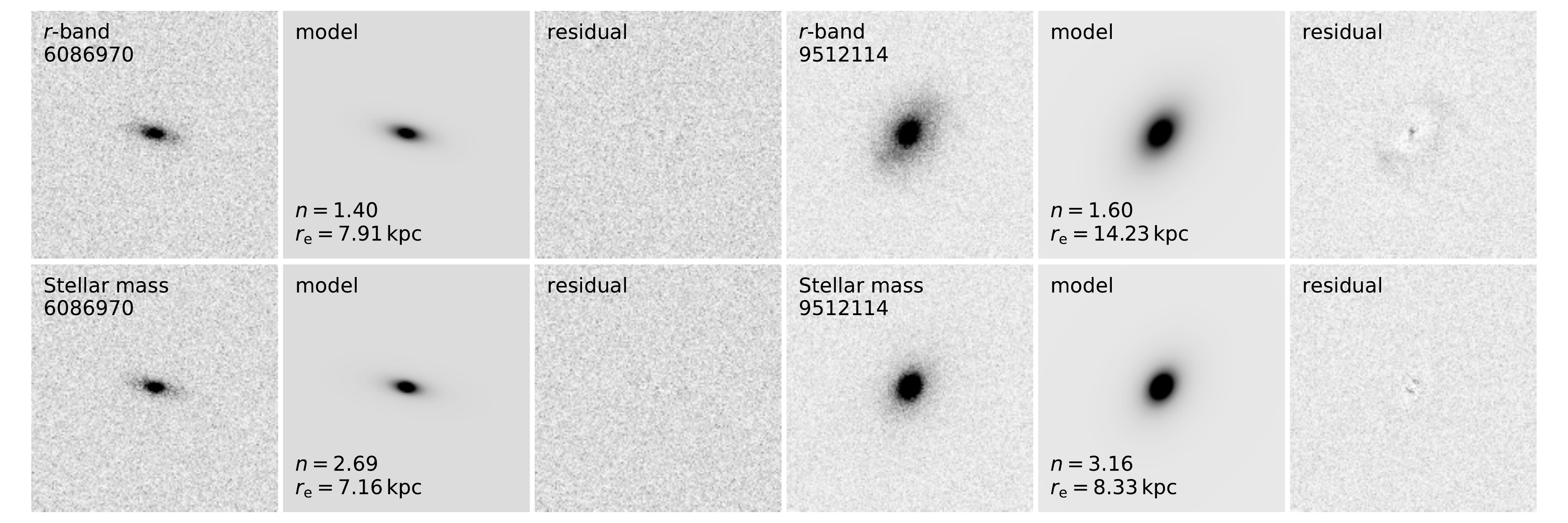}
    \caption{Results of the S\'ersic profile modelling with \textsc{Galfit} for the galaxies presented in Figs.~\ref{fig:example_skirt} and \ref{fig:example_massmap}. The top row shows the $r$-band image (left), best-fit S\'ersic model (middle), and residual (right) of the two galaxies, respectively. The bottom row shows the corresponding results for the stellar mass images.}
    \label{fig:example_galfit}
\end{figure*}

\subsubsection{Initial parameter estimation}\label{sec:sextractor}

As the S\'ersic model is described by five parameters, increased to seven free parameters by the addition of the galaxy centroid position, there is a vast parameter space to be explored to find their optimal values. It is therefore crucial to provide reasonable initial estimates of the S\'ersic parameters to reduce the computational cost, and avoid the fit to converge to a local, rather than global, minimum. 

We use \textsc{SExtractor} \citep{Bertin1996} to detect the source(s) present in each image and extract their photometric properties. Unlike real observations, the mock images include only mass and light from the vicinity of the galaxy potential minimum, and, in the majority of cases, there is thus only one source to be found by \textsc{SExtractor}. However, merging systems or small satellites of larger satellite galaxies may have been identified as a single galaxy by the \textsc{subfind} algorithm, but show two (or more) spatially distinct components in the imaging. We therefore run \textsc{SExtractor} with a setup akin to the `cold mode' employed by the \textsc{GALAPAGOS} code \citep[for details, see][]{Rix2004,Barden2012}, which was optimised to detect and deblend flux from bright sources. Specifically, we use a relatively high detection threshold, requiring $3\sigma$ detections over 15 adjacent pixels after smoothing with the default convolution kernel. To deblend the detected object(s), we use a number of 64 subthresholds (the levels between the detection threshold and maximum count value; $\rm DEBLEND\_NTHRESH = 64$) and a minimum contrast of $\rm DEBLEND\_MINCONT = 0.0001$. For each image, the corresponding sigma image (Section~\ref{sec:mock_ugriz}) is used to provide the algorithm with the root mean square (RMS) noise level, and the background is set to a fixed value of zero. We note that this procedure is vastly simplified in comparison with observational data, due to the fact that our images contain just one or few bright objects, and we have a perfect background subtraction and noise model. Our \textsc{SExtractor} results are therefore only weakly sensitive to changes in the parameters in the configuration file.

The output catalogue of \textsc{SExtractor} contains the centroid position (`X\_IMAGE', `Y\_IMAGE'), total flux (`FLUX\_AUTO'), half-light radius (`FLUX\_RADIUS'), ellipticity ($e\equiv 1-q$; `ELONGATION') and position angle ($\theta\equiv \phi + 90\degr$; `THETA\_IMAGE') of each extracted source. We use these to set the initial values for the position, $m$ or $M_{\rm *,Sersic}$, $q$, and $\phi$ of the S\'ersic model, respectively. For the initial value of $r_{\rm e,maj}$, we follow the approach by \citet{Kelvin2012} and correct the circularised radius from \textsc{SExtractor} to a major axis size, and account for the PSF convolution:
\begin{equation}
    r_{\rm e,maj} = \sqrt{ \frac{r_{\rm e,circ}^2}{q} - 0.32\,\Gamma^2 }\,,
\end{equation}
where $\Gamma$ is the FWHM of the PSF.
This leaves just one parameter, the S\'ersic index, which we set to an initial value of $n=4$.

\subsubsection{S\'ersic profile fitting}\label{sec:galfit}

We perform the S\'ersic modelling with \textsc{Galfit} \citep{Peng2010}, which uses the Levenberg-Marquardt algorithm to find the parameter values for which the total $\chi^2$ value of the image is minimised. 
To do so, the mock image, sigma (RMS) image and PSF (from Sections~\ref{sec:mock_ugriz} and \ref{sec:mock_im_mstar}) are provided as an input. We allow for multiple S\'ersic profiles in the configuration file, such that satellite galaxies (if present) are fit simultaneously with the primary galaxy, and initial parameters for each profile are set as described in the previous section. The sky background is fixed to a value of zero, although we investigate the effect of allowing for a variable sky component in the section~\ref{sec:sky_var}.

As a first pass, we do not place any constraints on the parameter values, to let the algorithm freely explore the parameter space. For most galaxies, this procedure leads to convergence with reasonable sizes and S\'ersic indices. Occasionally however, the S\'ersic index reaches implausible values (e.g., $n<0.2$), and we therefore rerun the fits for these objects with an additional constraint of $0.2<n<8.0$ for the primary component only, which can lead to convergence within this range. In Fig.~\ref{fig:example_galfit}, we demonstrate the S\'ersic modelling for the galaxies in shown in Figs.~\ref{fig:example_skirt} and \ref{fig:example_massmap}.

\subsubsection{Flags}\label{sec:flags}

We assess the quality of the fits by three criteria, which translate into a single combined flag: any fit that has converged at the boundary of the allowed range in $n$ is assigned a flag value of 1; a value of 2 is added to indicate images in which multiple components are fit simultaneously; a value of 4 is added to objects with bad fits, and is assigned on the basis of a visual inspection of the fits and residual images. This latter category consists of a mixture of objects, such as ongoing mergers that are simply not well described by S\'ersic profiles, brightest cluster galaxies that have highly complex morphologies, or simply failed fits that are unrealistically large in size. In few cases (27), we find that the \textsc{SExtractor}-detected sources are overdeblended, due to strong dust lanes or star-forming clumps in the disc being detected as separate objects. This only affects the optical images, and for these few galaxies we redo the \textsc{Galfit} fitting with a single component. 

We provide the final catalogues of the best-fit S\'ersic model parameters and flag values in Tables~\ref{tab:rband} and \ref{tab:mass}.
For the results presented in the following sections, we filter out all galaxies that contain a flag value of 1 (18 $r$-band fits, 33 stellar mass fits), as these measurements of the S\'ersic index and size are not robust. Galaxies with bad fits are also removed from the sample. Clearly, the definition of a `bad' fit is subjective, however, $<1\%$ of galaxies fall in this category (25 $r$-band fits, 10 stellar mass fits), and the population statistics are therefore likely unaffected (even at $M_*\gtrsim 10^{11.2}\,{\rm M_\odot}$\,, $<20\%$ of galaxies are excluded). With these quality criteria applied, 3560 galaxies remain with good fits in both the $r$-band and stellar mass imaging.

\begin{table*}
 \caption{Best-fit $r$-band structural parameters and uncertainties from \textsc{Galfit}. This table is available in its entirety online. }
 \label{tab:rband}
 \begin{tabular}{lcccccc}
  \hline
  GalaxyID & $m$ & $r_{\rm e}$ & $n$ & $q$ & $\phi$ & flag \\
   & mag & kpc & & & deg & \\
  \hline
  2 & $17.24 \pm 0.00$ & $2.77 \pm 0.02$ & $1.02 \pm 0.02$ & $0.44 \pm 0.00$ & $33.68 \pm 0.35$ & 0 \\
  13632 & $17.13 \pm 0.01$ & $3.02\pm0.05$ & $2.45\pm0.07$ & $0.74\pm0.01$ & $-39.91\pm1.20$ & 0 \\
  21794 & $17.69 \pm 0.01$ & $4.10\pm 0.06$ & $1.26\pm0.03$ & $0.50\pm0.01$ & $10.58\pm0.65$ & 0\\
  23302 & $17.47 \pm 0.01$ & $1.47\pm0.02$ & $3.50\pm0.18$ & $0.58\pm0.01$ & $55.00\pm1.03$ & 0 \\
  24478 & $18.19\pm0.01$ & $4.78\pm0.10$ & $1.01\pm0.04$ & $0.43\pm0.01$ & $62.11\pm0.75$ & 0 \\
  \hline
 \end{tabular}
\end{table*}

\begin{table*}
 \caption{Best-fit stellar mass structural parameters and uncertainties from \textsc{Galfit}. This table is available in its entirety online.}
 \label{tab:mass}
 \begin{tabular}{lcccccc}
  \hline
  GalaxyID & $\log(M_*/{\rm M_\odot})$ & $r_{\rm e}$ & $n$ & $q$ & $\phi$ & flag \\
   &  & kpc & & & deg & \\
  \hline
  2 & $10.837\pm0.000$ & $2.61\pm0.02$ & $0.88\pm0.02$ & $0.42\pm0.00$ & $34.02\pm0.24$ & 0 \\
  13632 & $10.848\pm 0.004$ & $2.63\pm0.02$ & $2.60\pm0.06$ & $0.77\pm0.01$ & $-40.93\pm1.07$ & 0 \\
  21794 & $10.592 \pm 0.004$ & $3.96\pm 0.05$ & $1.22\pm0.03$ & $0.49\pm0.00$ & $9.91\pm0.49$ & 0\\
  23302 & $10.647 \pm 0.000$ & $1.33\pm0.02$ & $2.72\pm0.12$ & $0.58\pm0.01$ & $52.99\pm0.84$ & 0 \\
  24478 & $10.383\pm0.004$ & $4.72\pm0.07$ & $0.92\pm0.03$ & $0.41\pm0.01$ & $63.50\pm0.54$ & 0 \\  
  \hline
 \end{tabular}
\end{table*}

\subsubsection{Measurement uncertainties}\label{sec:galfit_errors}

Although \textsc{Galfit} provides an estimated uncertainty on the measured structural parameters (limited to two decimal places), these tend to underestimate the true uncertainties \citep[for discussion, see][]{vdWel2012}. To obtain an estimate of the typical uncertainty on the different parameters, we create a second random noise realisation of the mock images, and repeat the S\'ersic profile fitting for this set of images.
By comparing the differences in the structural parameters between the two runs, we find that the scatter in $r_{\rm e,maj}$ corresponds to a typical uncertainty of $\delta\log(r_{\rm e,maj})=0.03\,$dex. Similarly, for the total magnitude and stellar mass $\delta\log(m_{\rm S{\acute e}rsic})=0.05\,$mag and $\delta\log(M_{\rm *,S{\acute e}rsic})=0.02\,$dex, respectively, and the S\'ersic index is the hardest to constrain precisely, with $\delta\log(n)=0.04\,$dex. These values are broadly consistent with the uncertainties found by \citet{vdWel2014a}, given that the typical signal-to-noise ratio of our images $\rm S/N \approx 100$ (where the S/N is calculated using the pixels belonging to the galaxy as identified by \textsc{SExtractor}).
Galaxies for which the \textsc{Galfit} estimates of the uncertainties are smaller than these values, are assigned the above typical values where needed.

\subsubsection{Sky background estimation}\label{sec:sky_var}

\begin{figure}
    \centering
    \includegraphics[width=\linewidth]{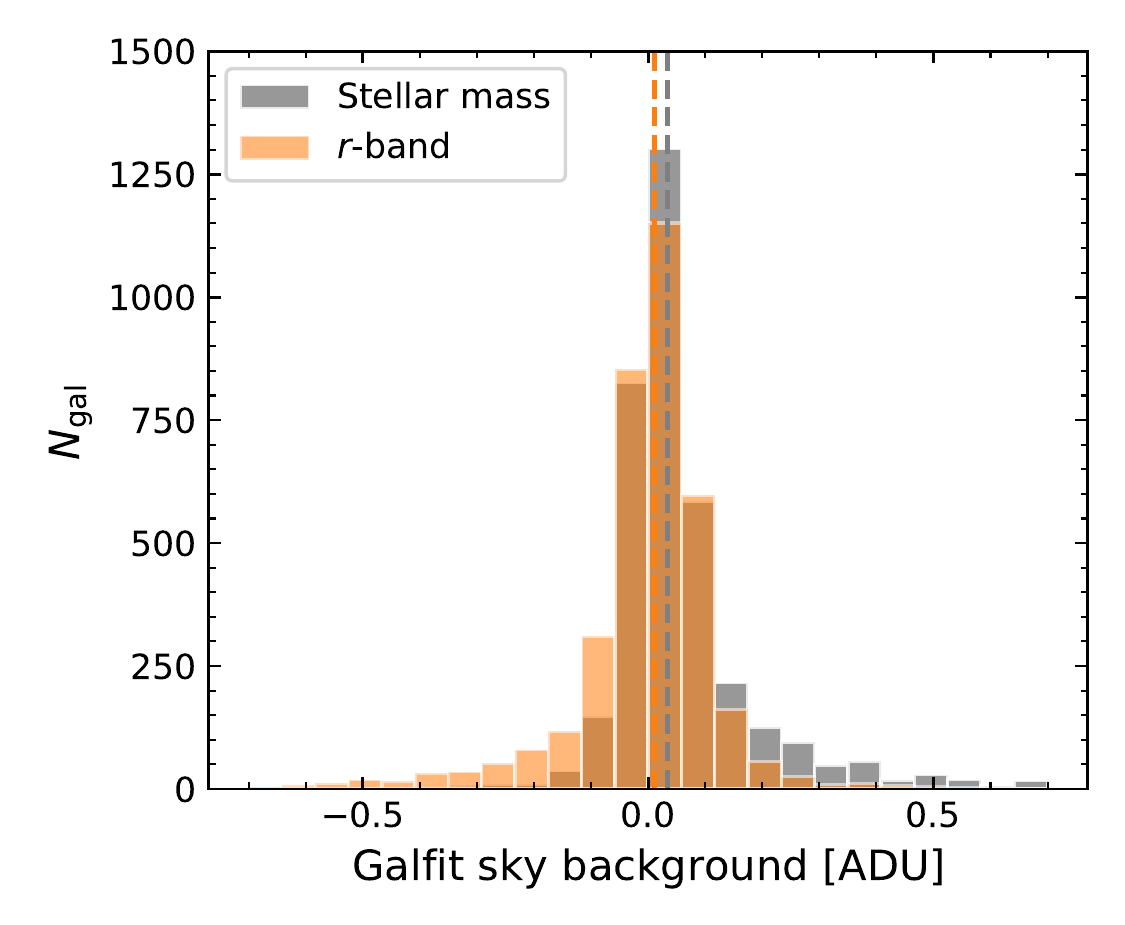}
    \caption{Distribution of the best-fit sky background level from \textsc{Galfit}. The median of the distribution (dashed line) is close to zero for both the $r$-band (orange) and stellar mass (grey) images with little scatter ($0.2$\,ADU; where the unit ADU is related to the number of photoelectrons as described in Section~\ref{sec:mock_ugriz}), which is negligible in comparison to the typical galaxy flux per pixel of order $\sim 10-10^2\,$ADU. We therefore also find no systematic differences between the best-fit S\'ersic models from the fits with a variable and fixed sky background.    }
    \label{fig:sky_vals}
\end{figure}

As noted in Section~\ref{sec:sextractor}, our images have perfect background subtraction by construction, which enables us to set the background to a fixed value of zero. However, obtaining an accurate background is often a challenge in observational studies, and commonly used tools such as \textsc{SExtractor} have been found to overestimate the sky background \citep{Haussler2007}. 

The source extraction, and in particular the S\'ersic modelling, is highly sensitive to the estimation of the background. To test whether the comparison we wish to make between our S\'ersic fits and those from observational data is affected by sky background uncertainties, we rerun both \textsc{SExtractor} and \textsc{Galfit} on the mock imaging with a variable sky component. We note that this test does not capture all the complexities faced in observational studies, where the background usually varies spatially across the image, but serves as a test for any systematic effects from including a nuisance parameter in the S\'ersic profile modelling.

From \textsc{SExtractor}, we obtain an initial estimate of the sky background in the image. We then create a sky component in the configuration file for \textsc{Galfit}, to be fitted simultaneously with the S\'ersic profile(s). An accurate fit of the background by \textsc{Galfit} requires a sufficiently large area of background pixels in comparison with the area spanned by the galaxy itself. With an image size of $60\arcsec$ on a side, this is the case for the majority of the sample. For very large galaxies this area is insufficient, causing the background to be overestimated due to confusion between the sky background and low surface brightness emission from the object itself. 

Fig.~\ref{fig:sky_vals} shows the distribution of the sky background as determined by \textsc{Galfit}, for both the $r$-band and stellar mass fits. Both distributions peak at a value of zero and show small scatter: the median of $0.01$\,ADU ($r$-band) or $0.03$\,ADU (stellar mass) and standard deviation of $0.2$\,ADU are well below the typical galaxy flux per pixel of order $\sim 10^2$\,ADU. For the stellar mass images, the asymmetric tail towards positive values of the sky background can be explained by the aforementioned effect of fitting the background in relatively small images. This effect is not present in the $r$-band images, as these images consist largely of empty background pixels (see Section~\ref{sec:mock_ugriz}). Most importantly, we find no systematic difference in the derived structural parameters between the fits with and without a variable sky component. The additional uncertainties on the structural parameters introduced by the variable sky component are also insignificant in comparison with the random uncertainties described in the previous section (\ref{sec:galfit_errors}): $\delta\log(r_{\rm e,maj})=0.006\,$dex, $\delta\log(m_{\rm S{\acute e}rsic})=0.013\,$mag or $\delta\log(M_{\rm *,S{\acute e}rsic})=0.005\,$dex, and $\delta\log(n)=0.009\,$dex.

\section{Galaxy sizes}\label{sec:results_sizes}

In this section we present the sizes measured with the S\'ersic modelling, and evaluate how the estimated half-mass and half-light radii differ from commonly used measures of size from the public EAGLE catalogues \citep{McAlpine2016}. In addition to the different measurement methods, we examine the effects of gradients in the stellar population properties and dust attenuation on the observed size. We then assess the impact of different size (and stellar mass) estimates on the obtained stellar mass-size relation, and compare with the observed mass-size relation at $z\sim0$.

\subsection{Do simulated galaxies follow S\'ersic profiles?}\label{sec:profiles}

\begin{figure*}
    \centering
    \includegraphics[width=\linewidth]{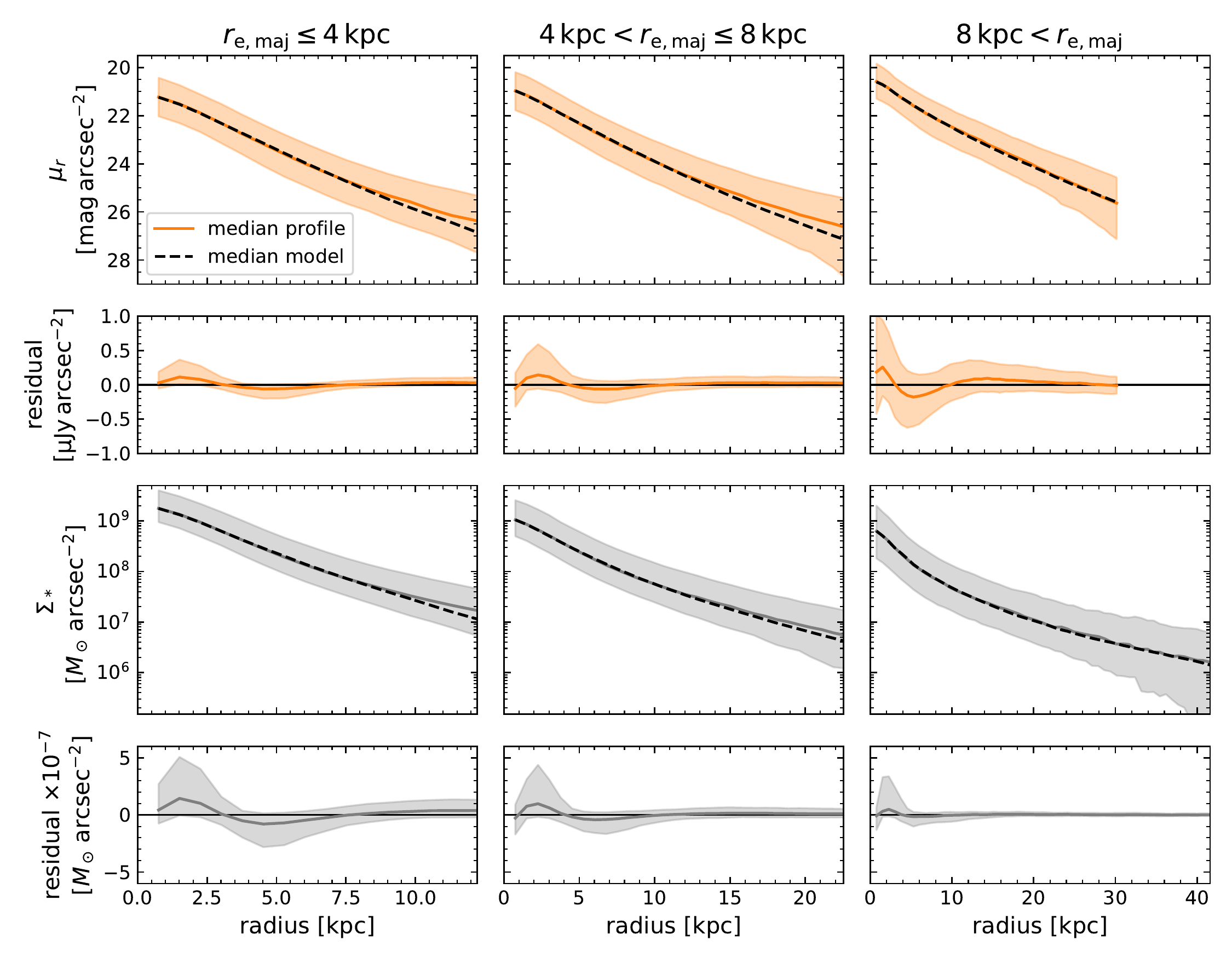}
    \caption{Surface brightness (orange) and stellar mass surface density (grey) profiles of EAGLE galaxies. The sample is divided into three bins of increasing half-light or half-mass radius, and the profiles are normalised to the median magnitude or stellar mass within each bin. Coloured lines and shaded areas in the first and third rows show the 16th, 50th, and 84th percentiles of the normalised profiles measured from the mock imaging, and dashed lines indicate the median of the normalised, best-fit S\'ersic models in each panel. Surface brightness profiles are cut off at $r=30\,$kpc, corresponding to the size of the \textsc{skirt} images. The median, 16th and 84th percentiles of the residuals, calculated as the difference between the normalised profiles and models, are shown in the second and fourth rows (in linear scale, as opposed to the logarithmic scale used for the profiles). The surface brightness and density profiles closely follow S\'ersic profiles, with only minor systematic features in the residuals.}
    \label{fig:sb_profs}
\end{figure*}

\begin{figure*}
    \centering
    \includegraphics[width=\linewidth]{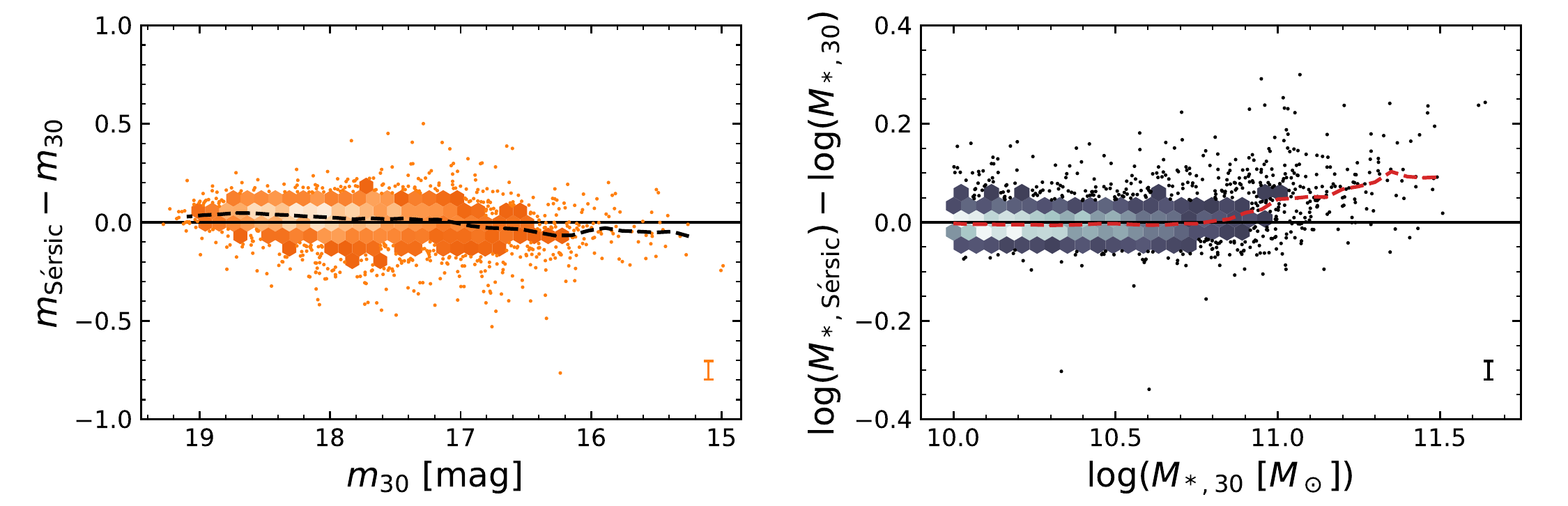}
    \caption{Recovered stellar mass and light with the S\'ersic modelling. The left-hand panel shows the difference between the total magnitude of the S\'ersic profile and the magnitude within a circular aperture of $30\,$kpc measured from the noise-free optical images, as a function of magnitude. The colour scale corresponds to the number density of data points, and individual galaxies are shown for sparsely populated regions \citep[created using \textsc{densityplot};][]{Krawczyk:13361}; the dashed line shows the running median. Similarly, the right-hand panel shows the deviation between the total stellar mass of the S\'ersic model and the conventional stellar mass of EAGLE galaxies (i.e., the total stellar particle mass within a spherical aperture of radius $30\,$kpc). On average, the models recover $98\%$ of the mass and light within $30\,$kpc. This increases toward higher masses and luminosities, where the $30\,$kpc aperture does not capture the full extent of the galaxy.  }

    \label{fig:flux_recovery}
\end{figure*}

However, before we make these different comparisons, we begin by asking whether the S\'ersic profile provides a good model for the surface brightness and density profiles of simulated galaxies. The simulation has finite resolution, set by both the mass of the particles and the gravitational softening scale of $\approx 2\,$kpc. 
Although \citet{Schaller2015} showed that the (3D) density profiles of the stellar and dark matter mass are on average well converged on scales $\gtrsim 2\,$kpc, for most galaxies the inner few kpc of the density profile will drive the fit of the S\'ersic profile, as this is where the majority of the high S/N flux is concentrated in the image. 

To gauge whether the finite resolution leads to systematic deviations from the S\'ersic model, we compare the S\'ersic profiles with the azimuthally-averaged profiles from the mock images. We first extract the surface brightness ($\mu_r$) and stellar mass surface density ($\Sigma_*$) profiles from the mock images and the best-fit models, by measuring the flux in elliptical apertures with the axis ratio and position angle from the best-fit S\'ersic model. As we may expect the resolution to have a different effect on the profiles depending on the galaxy size itself, the sample is divided in three bins according to the half-light or half-mass radius. 

In the upper panels of Fig.~\ref{fig:sb_profs}, we show the median, 16th and 84th percentiles of the observed $\mu_r$ profiles as a function of the half-light radius. For each size bin, the profiles are normalised to the median magnitude within the bin, and the scatter therefore represents a difference in the profile shape only. Underneath, we show the median, 16th and 84th percentiles of the residual profiles, which are calculated as the difference between the normalised profiles and best-fit models. For the largest size bin, the profiles are cut off at 30\,kpc, because of the limited spatial extent of the \textsc{skirt} images (see Section~\ref{sec:mock_ugriz}).
Similarly, the lower set of panels show the $\Sigma_*$ profiles as a function of the half-mass radius, normalised to the median stellar mass in each bin, as well as the residual profiles.

The simulated galaxies are generally well described by the S\'ersic models, as there are only 
some minor systematic features visible: in the left-hand panels (i.e., the smallest sizes), there is positive residual flux at $r\approx 2\,$kpc ($r\approx 0.6\,r_{\rm e,maj}$), whereas the region around $r\approx 5\,$kpc ($r\approx 1.7\,r_{\rm e,maj}$) is oversubtracted. At large radii, low surface brightness emission is also not fully captured by the single S\'ersic profile. In the larger size bins, similar residual features appear around the same absolute radii (and thus at a smaller number of effective radii), suggesting that the limited resolution of the simulations has a small, systematic effect on the profiles.

\begin{figure*}
    \centering
    \includegraphics[width=0.9\linewidth]{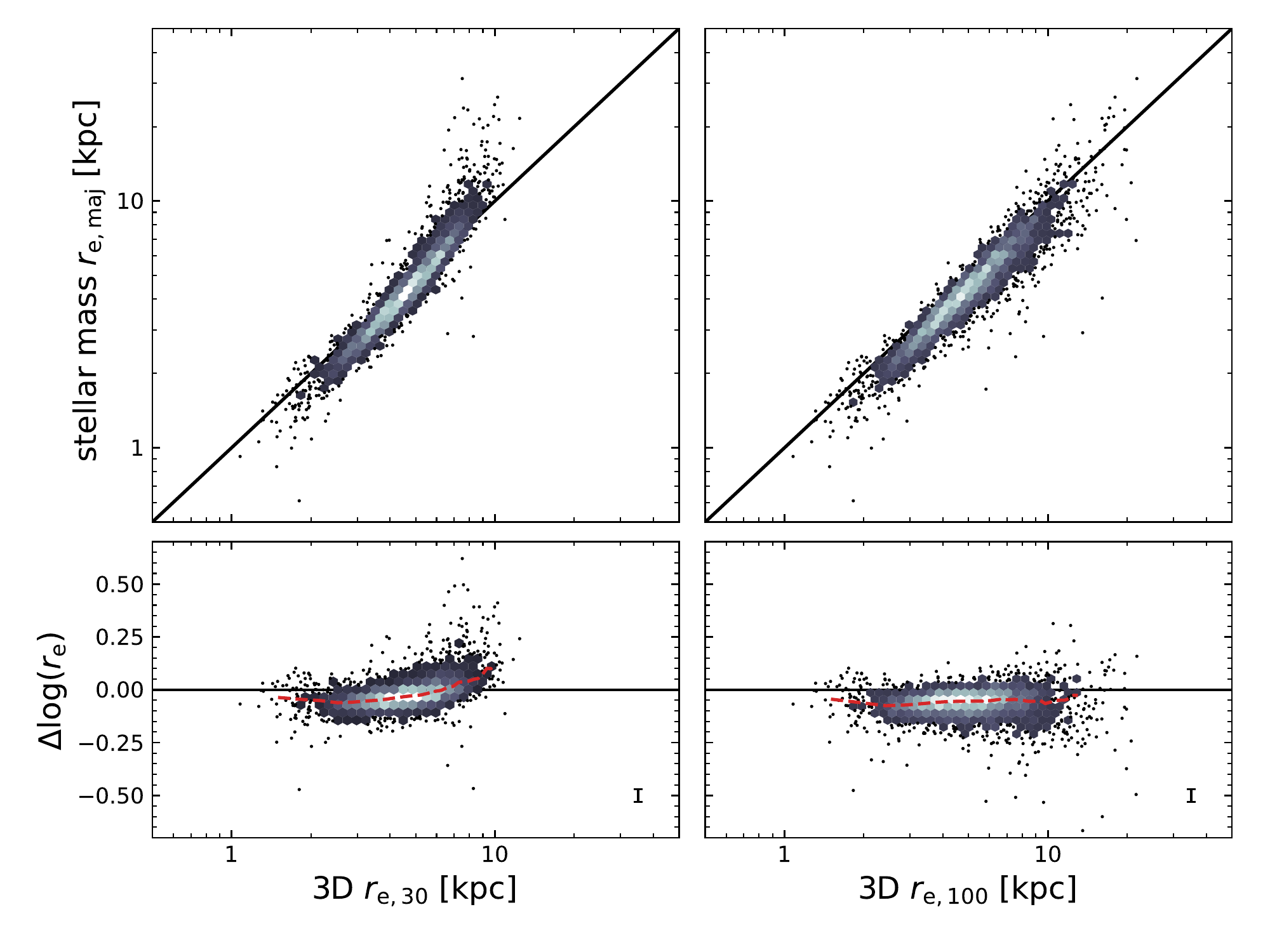}
    \caption{Comparison between different measures of the stellar half-mass radius. The upper panels show the semi-major axis of the best-fit S\'ersic model to the stellar mass imaging versus the 3D half-mass radius within a spherical aperture of radius $30\,$kpc (left) and $100\,$kpc (right). The lower panels show the difference between the sizes as a function of radius, together with the running median (dashed lines). There is a systematic discrepancy between the different measures of size, which depends on the radius itself: at small radii, the S\'ersic sizes are smaller by $\approx-0.05\,$dex (left) or $\approx-0.06\,$dex (right); this discrepancy decreases slightly toward larger radii in the right-hand panel, to a difference of $\approx-0.05\,$dex. The effect is significantly stronger in the left-hand panel, where the difference even changes sign, reflecting the fact that for massive galaxies the $30\,$kpc aperture underestimates the full extent of the galaxy. }
    \label{fig:3d_size_comp}
\end{figure*}

Furthermore, we find that the S\'ersic models perform well when comparing the integrated luminosity and stellar mass with the input data. Fig.~\ref{fig:flux_recovery} shows the difference between the total $r$-band magnitude ($m_{\rm S{\acute e}rsic}$) and the magnitude measured within $30\,$kpc in the \textsc{skirt} image (i.e., not including noise or instrumental effects). Similarly, the right-hand panel shows the difference in the total stellar mass ($M_{\rm *, S{\acute e}rsic}$) and the stellar mass within a $30\,$kpc aperture ($M_{\rm *,30}$) as a function of $M_{\rm *,30}$. Typically, $98\%$ of the luminosity or stellar mass is recovered in the S\'ersic fit. At high mass and high luminosity there is an increasingly stronger deviation, demonstrating that the $30\,$kpc aperture does not capture the full extent of the galaxy \citep[as noted previously by][]{Schaye2015}. In Appendix~\ref{apdx:aperture}, we also present the difference between $M_{\rm *, S{\acute e}rsic}$ and the stellar mass obtained for other aperture sizes (50\,kpc, 70\,kpc, 100\,kpc, and the entire subhalo mass), finding that $M_{\rm *, S{\acute e}rsic}$ is approximately equivalent to the stellar mass enclosed within a spherical aperture of radius 70\,kpc for very massive galaxies.

We therefore conclude that the S\'ersic model provides a good description of both the stellar mass surface density profiles and the surface brightness profiles of EAGLE galaxies, and defer a further discussion of the minor systematic residuals to Section~\ref{sec:discussion_profiles}.

\subsection{Comparing different measures of size}\label{sec:size_comp}

Size estimates in the EAGLE data release are based on a growth-curve method and come in two variations \citep[see also][]{Furlong2017}. The first method computes the total stellar mass belonging to a single subhalo within a spherical aperture of radius $R$ centred around the minimum of the potential, after which spherical apertures of increasing radius are constructed to find the radius that encloses $M_*(<R)/2$. From hereon, we will refer to this half-mass radius as the 3D $r_{\rm e, R}$. The second method also uses the total stellar mass within a spherical aperture of radius $R$ as starting point, however, the half-mass radius is now measured from a 2D projection of the stellar mass distribution: circular apertures of increasing radius are constructed to find the radius that encloses $M_*(<R)/2$. This computation is done for projections in three orthogonal planes, and the average of the three measurements then gives the 2D $R_{\rm e, R}$. With two different aperture sizes, $R=30\,$kpc and $R=100\,$kpc, there are four different estimates of the half-mass radius in total.

The difference between the 3D and 2D sizes is significant, with the former being on average a factor of $4/3$ larger, as is to be expected for spheroidal systems. More importantly, we find that this factor is not dependent on the galaxy mass or the sSFR. This is also apparent in Fig.~\ref{fig:axis_ratio} (discussed in Section~\ref{sec:axis_ratio}), which shows that the median projected axis ratio is approximately constant across the six bins in stellar mass and sSFR. To compare with the S\'ersic profile sizes, we can thus focus on just one of the two methods described above. In what follows, the results then translate to the other measure by a constant factor.

As the 2D $R_{\rm e, R}$ is by definition a circularised quantity, which differs systematically from the semi-major axis of the S\'ersic profile by a factor $\sqrt{q}$, we choose to use the 3D $r_{\rm e, R}$ for our comparison. Fig.~\ref{fig:3d_size_comp} shows the stellar half-mass radius from the \textsc{Galfit} modelling as a function of the 3D $r_{\rm e, 30}$ (left) and $r_{\rm e, 100}$ (right). The bottom panels additionally show the difference between the two size estimates ($\log(r_{\rm e,maj}/r_{\rm e,R})$), together with the running median (dashed lines).

There are small, but significant, systematic differences between the 3D and S\'ersic sizes. For small galaxies ($r_{\rm e}\lesssim4\,$kpc), the major axis sizes of the S\'ersic fits are smaller by a constant factor of approximately $0.89\,r_{\rm e, 30}$ and $0.86\,r_{\rm e, 100}$ (or equivalently, a mean difference of $-0.053\,$dex and $-0.066\,$dex, respectively). However, there is a dependence on radius, particularly in the left-hand panel, where at large radii the S\'ersic-derived half-mass radii are systematically larger. As also discussed in Section~\ref{sec:profiles}, for very massive galaxies the spherical aperture of $30\,$kpc is simply too small to encompass the full extent of the galaxy, and the 3D half-mass radii are therefore underestimated. For the larger aperture of $100\,$kpc this effect is greatly reduced, although there is still a slight increase in $\Delta\log(r_{\rm e})$ with increasing radius, with a mean difference of $-0.053\,$dex for galaxies with $r_{\rm e,100}>4\,$kpc. The size discrepancies found here appear to be slightly larger than the predictions by \citet{vdVen2021} (of $\Delta\log(r_{\rm e})\approx0.02\,$dex), who derived an analytical prescription for the conversion from S\'ersic profile sizes to 3D sizes. On the other hand, the two results are likely to be consistent when taking into account the fact that the axis ratio distributions differ systematically between EAGLE and local observations (as oblate systems in EAGLE are not sufficiently flattened, see Section~\ref{sec:axis_ratio}).

\begin{figure}
    \centering
    \includegraphics[width=\linewidth]{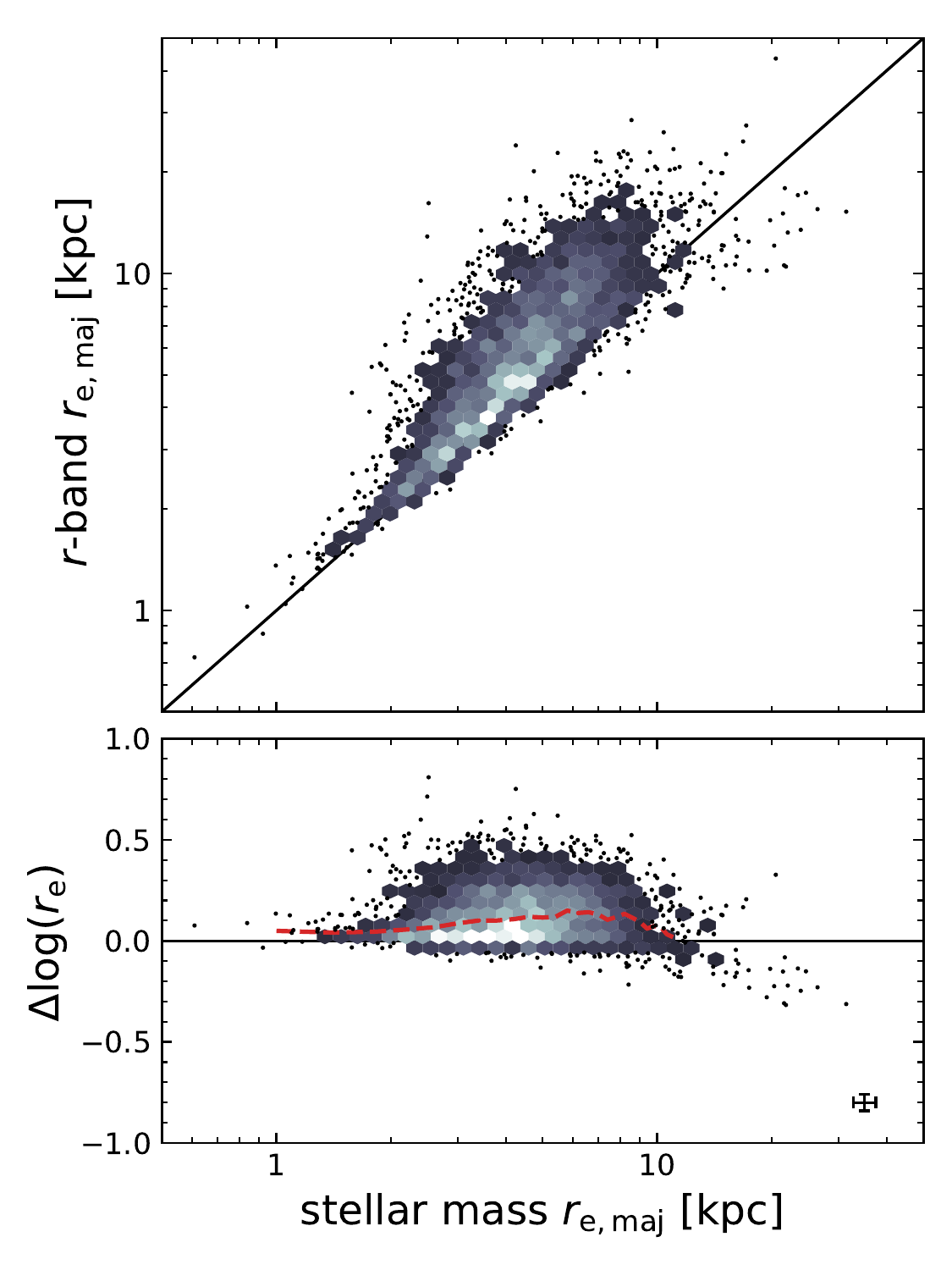}
    \caption{Comparison between the sizes obtained from the S\'ersic profile fitting to the $r$-band and stellar mass image data. The dashed line shows the running median of the logarithmic difference between the two size estimates. Both quantities were estimated with the same methodology and using image data with similar noise and equal spatial resolution, and discrepancies can therefore be attributed entirely to radial variations in the mass-to-light ratio within galaxies. The half-light radii are systematically larger than the half-mass radii (typically, 25\% larger), and this discrepancy increases slightly toward larger radii, albeit with large scatter. }
    \label{fig:mass_vs_light}
\end{figure}

Thus far, however, we have only compared stellar half-mass radii, which give insight into the effect of different methodologies. This does not account for the effects of dust and stellar population gradients that affect the shapes of the light profiles, and hence also the inferred sizes. In Fig.~\ref{fig:mass_vs_light}, we show how the $r$-band half-light radii compare with the stellar half-mass radii. Here, the methodology is the same for both axes, and discrepancies are therefore entirely due to radial variations in the mass-to-light ratio ($M_*/L_r$).

The half-light radii are systematically larger than the half-mass radii, with an offset that depends weakly on radius. The two are comparable only for small galaxies ($r_{\rm e,maj}\lesssim2\,$kpc), which are mainly compact quiescent galaxies that may be expected to have only weak $M_*/L_r$ gradients, although we caution that these galaxies are smaller than the PSF FWHM ($2.6\,$kpc). The bottom panel shows the size difference as a function of the half-mass radius, together with the running median (dashed line): on average, the $r$-band radii are 40\% larger ($0.14\,$dex), with a median of 25\% ($0.10\,$dex). However, there is also significant scatter (of $0.13\,$dex), which is asymmetric with excesses of up to $\sim 0.5\,$dex: gradients in $M_*/L_r$ can thus have a great effect on the inferred size for individual galaxies. 

\begin{figure*}
    \centering
    \includegraphics[width=0.95\linewidth]{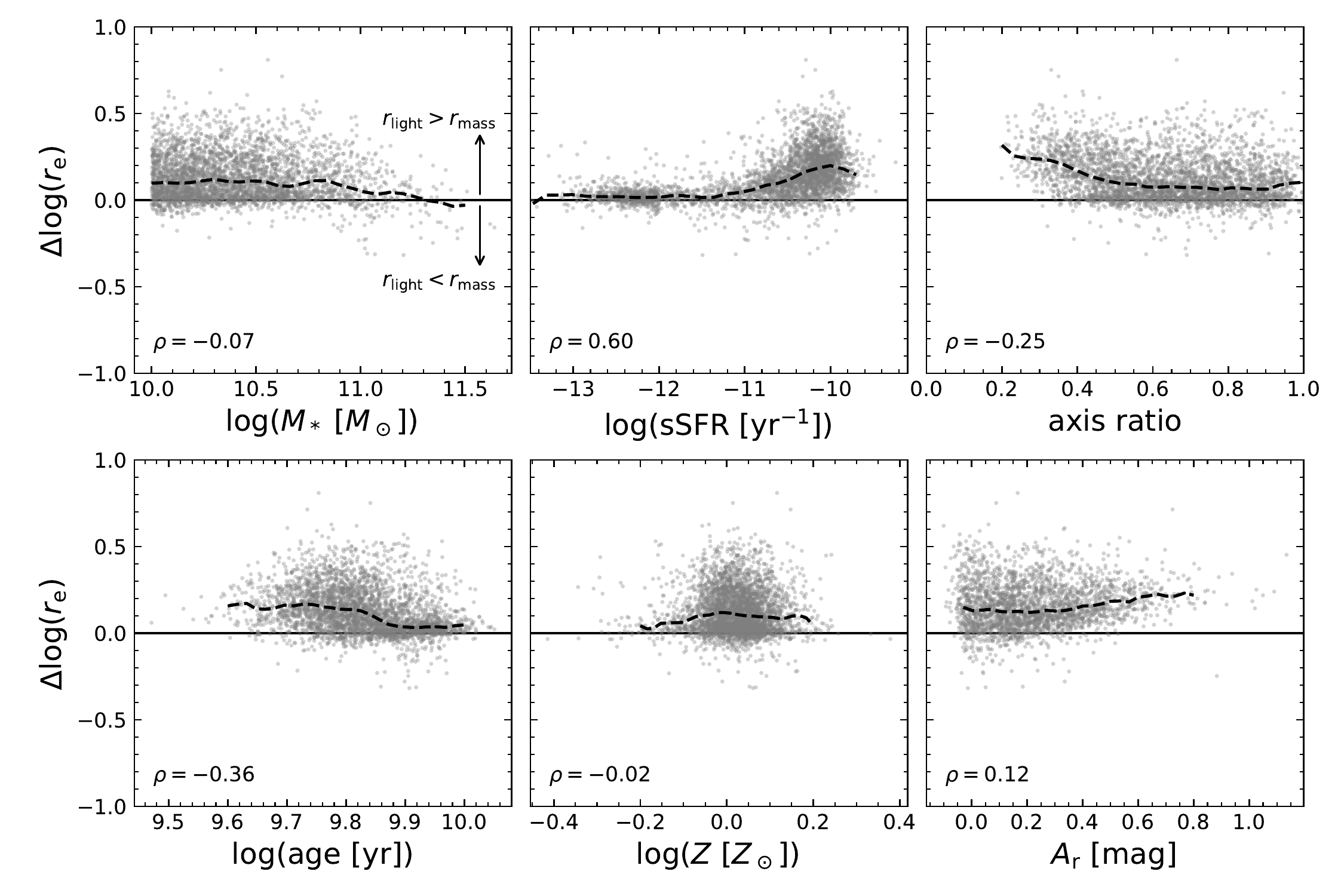}
    \caption{Difference between the $r$-band half-light radii and stellar half-mass radii ($\Delta\log(r_{\rm e})$) as a function of stellar population and dust properties. Dashed lines show the running median in each panel, and the value of the Spearman rank correlation coefficient ($\rho$) is indicated in each panel. The size difference is largely independent of the stellar mass and mass-weighted stellar metallicity, but depends strongly on the star formation activity, reflected by the positive correlation with the sSFR and negative correlation with the mass-weighted stellar age. Dust also has a significant effect, as edge-on galaxies show a stronger size discrepancy, and the $r$-band dust attenuation ($A_{\rm r}$) correlates weakly with $\Delta\log(r_{\rm e})$. Star formation in the outskirts of galaxies, as well as dust attenuation in the central regions therefore likely drive the discrepancies between light and mass-weighted sizes.  }
    \label{fig:delta_re}
\end{figure*}

Fig.~\ref{fig:delta_re} examines the origins of the size differences, by showing $\Delta\log(r_{\rm e})$ as a function of different galaxy properties. In addition to the stellar mass and specific star formation rate (sSFR) measured within a spherical aperture of radius 30\,kpc, we extract the mean mass-weighted age and metallicity of the stellar particles within the same aperture. We note that, for visualisation purposes only, we have added a value of $0.01\,{\rm M_\odot}\,\rm yr^{-1}$ to the instantaneous SFR before calculating the sSFR. Moreover, we estimate the dust attenuation in the $r$-band ($A_{\rm r}$) by calculating the difference between the rest-frame absolute magnitudes with and without dust from \citet{Trayford2015,Trayford2017}, although we note that the attenuated magnitudes are only available for galaxies with a minimum of 250 dust particles (2590 galaxies).

The size discrepancy is independent of the stellar mass below $M_*\sim10^{11}\,{\rm M_\odot}$, but at the high mass end the half-light radii become comparable to the half-mass radii. This may reflect minimal dust attenuation or colour gradients at these high masses, although this may also be partially due to the limited spatial extent of the original \textsc{skirt} images (60\,kpc versus 114\,kpc in the stellar mass images).

On the other hand, there is a strong correlation with the sSFR, and a similar trend is visible for the stellar age, with the youngest galaxies having much higher values of $\Delta\log(r_{\rm e})$ (by $\approx0.15\,$dex) than the very oldest systems. Interestingly, we find no such correlation with the stellar metallicity.

Furthermore, edge-on galaxies (low projected axis ratios), which tend to have higher optical depths due to dust, have relatively large half-light radii. Observationally, this effect may be even stronger, as edge-on EAGLE galaxies are thicker than observed in the local Universe and thus also have significantly lower dust optical depths \citep[see][]{Trayford2017}. The effects of dust are also visible in the lower right panel, which shows a weak, positive correlation between $\Delta\log(r_{\rm e})$ and the $r$-band dust attenuation. If we select only the highly star-forming galaxies ($\rm sSFR>10^{-10.4}\,yr^{-1}$), the effects of dust become even more pronounced: although the correlation with $A_{\rm r}$ becomes negligible, the anti-correlation with the axis ratio becomes slightly stronger (Spearman rank coefficient $\rho=-0.32$), which suggests that the dust geometry is an important factor. The increased spatial extent in the $r$-band imaging with respect to the stellar mass imaging can therefore be attributed to the presence of bright star-forming regions in the outskirts of galaxies and/or significant dust attenuation in the centre.

\subsection{Stellar mass-size relation}\label{sec:mass_size}

\begin{figure*}
    \centering
    \includegraphics[width=\linewidth]{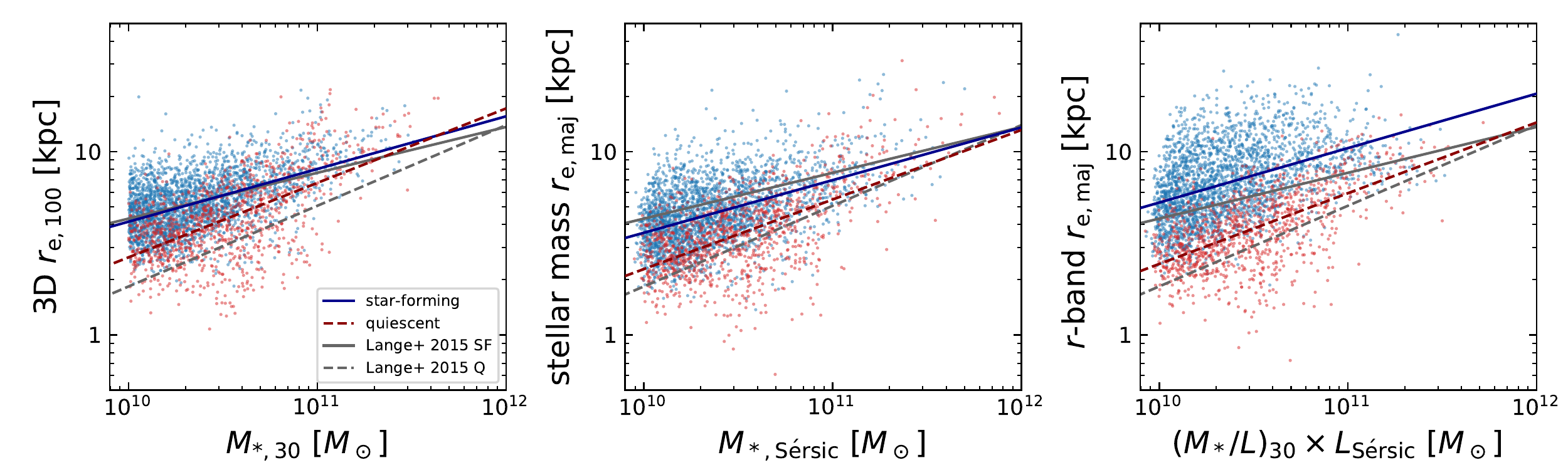}
    \caption{Stellar mass-size relation of quiescent (red) and star-forming (blue) EAGLE galaxies for three different measures of galaxy size and stellar mass. The left-hand panel shows the stellar masses and half-mass radii from the public EAGLE catalogues, which are measurements within spherical apertures of fixed radius. The middle panel shows the half-mass radius and total mass of the best-fit S\'ersic profiles to the stellar mass imaging. On the right, the results from the $r$-band S\'ersic fits are presented, with stellar masses corrected to the total luminosity of the best-fit profile. Coloured lines show the best-fit power law relations in each panel. Grey lines indicate the best-fit $r$-band mass-size relations from \citet{Lange2015}, for star-forming (solid) and quiescent (dashed) galaxies at $z\sim0$ in the GAMA survey (where quiescence is defined by the dust-corrected, rest-frame $u-r$ colour). Only in the right panel, with the fully forward-modelled sizes, is there a clear separation between the star-forming and quiescent populations, and are both the slope and scatter about the relation comparable with observations (see main text). The zero-point offsets are slightly higher than in GAMA, however, indicating that both quiescent and star-forming galaxies in EAGLE are systematically larger (by $\approx 0.1\,$dex) than observed in the local universe.}
    \label{fig:mass_size}
\end{figure*}

As described in \citet{Crain2015}, for the EAGLE simulations the stellar mass-size relation of late-type galaxies at $z\sim0$ was used in the calibration of the subgrid model parameters. Specifically, of the four subgrid models considered, three were rejected due to the simulations producing unrealistic size distributions for the massive galaxy population \citep[$>0.2\,$dex below the mass-size relation from][]{Shen2003}, and mass-size relations that decline with mass, rather than increase. Although the subgrid model was not fine-tuned to reproduce the observed mass-size relation, the low-redshift mass-size relation in EAGLE can also not be considered to be a true prediction of the simulation \citep{Schaye2015}.
However, this calibration was done using stellar half-mass radii that were measured by fitting S\'ersic profiles to the projected, azimuthally-averaged stellar mass density profiles, and used only the subset of EAGLE galaxies with S\'ersic index $n<2.5$. We showed previously that there are significant, systematic differences between different measures of size, which may therefore affect the inferred mass-size relation, and potentially also the calibration of a subgrid model.

We explore the effects of different size estimates on the mass-size relation in Fig.~\ref{fig:mass_size}. The left-hand panel shows results that are similar to the work by \citet{Furlong2017}, who presented the redshift evolution of the mass-size relation in the EAGLE simulations. They defined the galaxy stellar mass as the mass enclosed within a spherical aperture of radius $30\,$kpc, and the stellar half-mass radius as the 3D $r_{\rm e,100}$ (see also Section~\ref{sec:size_comp}). Observations indicate different evolution for late- and early-type galaxies, which holds true regardless of the method used to define `late' versus `early' \citep[by colour, morphology, or SFR; see, e.g.,][]{Shen2003}. We therefore divide our sample by the instantaneous sSFR within the 30\,kpc spherical aperture, and define the late-type or star-forming population as having $\rm sSFR > 10^{-11}\,yr^{-1}$; the star-forming and quiescent galaxies in Fig.~\ref{fig:mass_size} are indicated in blue and red, respectively.

Perhaps unsurprisingly, considering the subgrid model calibration, the star-forming EAGLE galaxies closely follow the observed mass-size relation at $z\sim0$ measured by \citet{Lange2015} using $r$-band S\'ersic models of GAMA galaxies. The grey solid and dashed lines show the single power-law fits ($r_{\rm e} = a\,(M_*/{\rm M_\odot})^b$) to the star-forming and quiescent subsamples, respectively, where quiescence for GAMA galaxies is defined using the rest-frame $u-r$ colour that is corrected for dust attenuation within the galaxy. 
If we fix the exponent to the local relation ($b=0.25\pm0.02$), and perform a least-squares fit in logarithmic space to determine the normalisation $a$, we find excellent agreement between EAGLE ($\log(a)=-1.870\pm0.003$; where the error bar is obtained via bootstrap resampling) and GAMA ($\log(a)=-1.87\pm0.05$). If we instead fit both $a$ and $b$ simultaneously (coloured lines), we find an exponent $b=0.287\pm0.009$ that is slightly steeper than observed, however, in good agreement with observations when considering the stellar mass limit imposed here ($M_*>10^{10}\,\rm M_\odot$ versus $M_*\gtrsim10^{9}\,\rm M_\odot$ for GAMA) and the fact that the mass-size relation has been shown to steepen toward high stellar mass \citep{Shen2003}.

The quiescent population, on the other hand, deviates strongly from the observed relation (dashed lines). Although the best-fit exponent of $b=0.406\pm0.017$ is close to the observed value of $b=0.44\pm0.02$, the normalisation is significantly higher: at fixed $b=0.44$, the EAGLE galaxies are $0.14\pm0.03$\,dex larger than observed ($\log(a) = -3.994\pm0.005$ versus $\log(a) = -4.14\pm0.03$).

However, the relations from \citet{Lange2015} are based on semi-major axis sizes from S\'ersic models. The middle panel of Fig.~\ref{fig:mass_size} shows the result of using the half-mass radii obtained with the S\'ersic profile fits for the stellar mass images. We emphasise that not only the size changes with respect to the left-hand panel, but also the stellar mass is replaced with the total mass of the best-fit S\'ersic profile. 

To take into account measurement uncertainties and the intrinsic scatter about the relation, we follow the maximum likelihood fitting method described by \citet{vdWel2014a} to estimate the best-fit parameters of the power-law model. This method assumes there is intrinsic scatter (i.e., not due to measurement uncertainties) about the mass-size relation that follows a Gaussian distribution, and fits the intrinsic scatter as an additional variable to the zero point ($a$) and slope ($b$). Moreover, uncertainties in $M_*$ are treated as an additional uncertainty in $\log(r_{\rm e})$.

We find that the slope of the relation for the S\'ersic model sizes is changed minimally with respect to the aperture-based sizes, with $b=0.287\pm0.010$ and $b=0.379\pm0.016$ for the star-forming and quiescent samples, respectively. 
There is a significant change in the intercept, however, as this deviates by $-0.060\pm0.004\,$dex and $-0.077\pm0.008\,$dex respectively for the star-forming and quiescent populations. These values are in line with the systematic offsets found in Fig.~\ref{fig:3d_size_comp}, and slightly enhanced by the fact that the stellar masses are also marginally smaller than the aperture-derived masses for the majority of the sample (Fig.~\ref{fig:flux_recovery}).
The result of moving from 3D half-mass sizes to major axis sizes from the stellar mass S\'ersic models is thus that the star-forming population appears systematically smaller than the observed mass-size relation by 0.06\,dex. On the other hand, the agreement with observations is significantly improved for the quiescent population, although these galaxies are still systematically larger than observed by 0.06\,dex.

Lastly, we take into account the effects of stellar population gradients and dust, by using the S\'ersic fits to the $r$-band imaging rather than the stellar mass fits. Again, it is not only the parameter on the vertical axis that changes, but we also adjust the stellar mass: we correct the aperture-based mass ($M_{*,30}$) by multiplying with the ratio of the total flux of the S\'ersic profile and the flux measured within a circular aperture of $30\,$kpc.

The resulting mass-size relation in the right-hand panel differs from the other two panels by not just the zero points offsets, but also the scatter. The first effect is mainly in the relative difference between the star-forming and quiescent populations. Whereas the two populations overlap quite significantly when considering the half-mass radii, there is a larger separation when using the half-light radii. Interestingly, it is the star-forming population that changes with respect to the middle panel: the quiescent population is moved only slightly, as these sizes are larger than the observed relation by $0.10\pm0.03$\,dex (at fixed $b=0.44$, $\log(a)=-4.040\pm0.005$). If we also fit the exponent, we find $b=0.386\pm0.015$, which is slightly shallower than the observed value (by $2.2\sigma$), although this measurement may be affected by the limited spatial extent of the $r$-band images (as discussed in Section~\ref{sec:size_comp}). On the other hand, the star-forming population moves towards much larger $r_{\rm e}$ at fixed stellar mass, and is $0.11\pm0.05\,$dex larger than the observed relation ($\log(a)=-1.760\pm0.003$ at fixed $b=0.25$). The best-fit exponent, $b=0.297\pm0.012$, is slightly steeper than observed (by $1.9\sigma$), but likely in good agreement with observations when taking into account the difference in the mass range used for the fitting.

Both populations are thus $\approx0.1\,$dex larger at fixed stellar mass than observed, but the separation between the quiescent and star-forming populations matches that of the observed relations almost exactly.
We can therefore conclude that colour gradients strongly affect the mass-size relation of the star-forming sample, and have only a moderate effect on the quiescent population.

The second difference with respect to the other panels is in the scatter in $\log(r_{\rm e})$ about the relation. For the star-forming population, the scatter in the half-light radii appears to be much closer to the observed scatter: using the normalised median absolute deviation (NMAD), we find $\sigma(\log\,r_{\rm e}) = 0.19$\,dex for the half-light radii, versus $\sigma(\log\,r_{\rm e}) = 0.14\,$dex and $\sigma(\log\,r_{\rm e}) = 0.15\,$dex for the 3D and S\'ersic half-mass radii, respectively. On the other hand, the scatter for the quiescent population is approximately equal for all three measures of size (from left to right, $\sigma(\log\,r_{\rm e}) = 0.16\,$dex, $\sigma(\log\,r_{\rm e}) = 0.15\,$dex and $\sigma(\log\,r_{\rm e}) = 0.15\,$dex). Although these measurements are not provided explicitly by \citet{Lange2015}, we obtain $\sigma(\log\,r_{\rm e}) = 0.20$ (star-forming) and $\sigma(\log\,r_{\rm e}) = 0.18$ (quiescent) for a $z\sim0.1$ comparison sample selected from GAMA (sample selection described in Section~\ref{sec:results_sindex}). We note that the uncertainties on the size measurements in GAMA are expected to be larger than is the case for the EAGLE galaxies (e.g., due to additional uncertainties from the sky background). Therefore, whereas the observed scatter about the mass-size relation of star-forming galaxies agrees well between EAGLE and GAMA, the intrinsic scatter may be slightly too large for the EAGLE galaxies.

\section{Galaxy morphologies}\label{sec:results_morph}

We now turn to the morphological properties of the EAGLE galaxies as quantified by the S\'ersic index and projected axis ratio. We compare our results with a low-redshift sample of galaxies selected from the GAMA survey, which is approximately volume-limited above our stellar mass limit at $z\sim0.1$. The optical imaging and derived data products of GAMA are largely based on SDSS imaging, and are therefore of similar quality to our constructed mock images and model fits.

\subsection{S\'ersic indices}\label{sec:results_sindex}

\begin{figure*}
    \centering
    \includegraphics[width=0.97\linewidth]{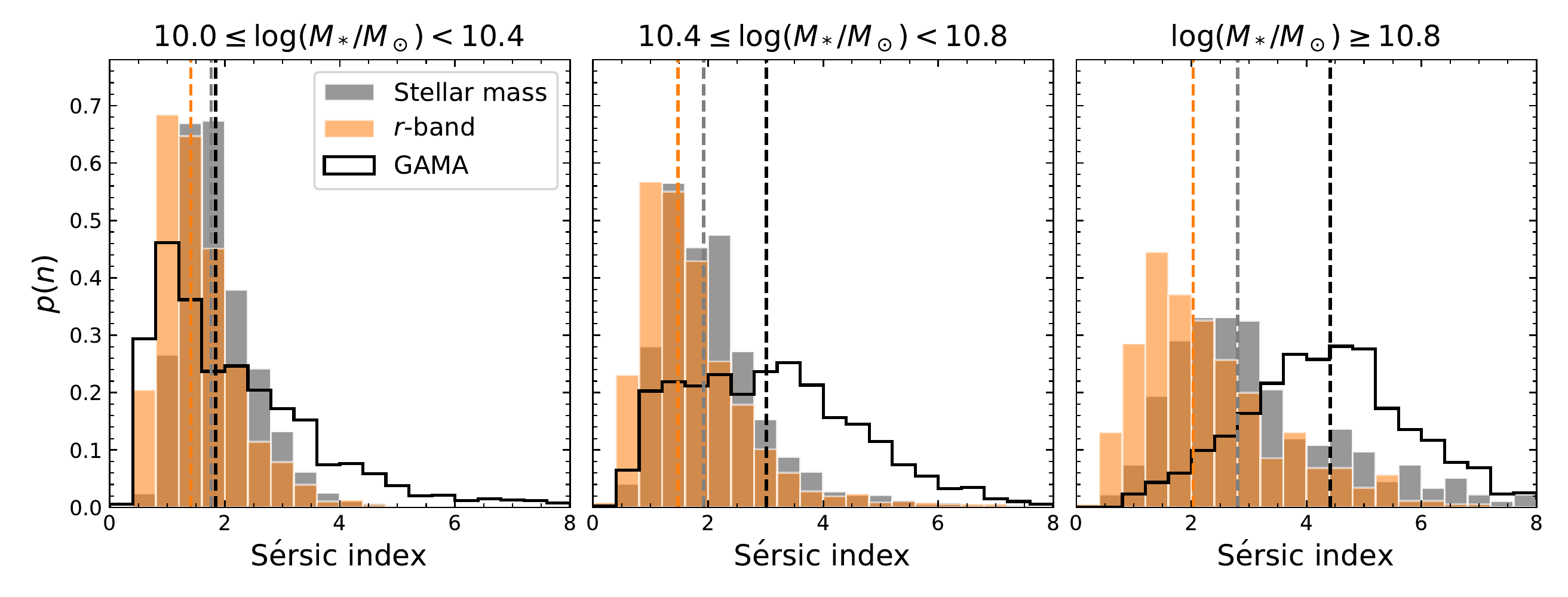}
    \caption{S\'ersic index distributions of EAGLE galaxies at $z=0.1$, in bins of increasing stellar mass. Results of the fits to the stellar mass and $r$-band images are shown in grey and orange, respectively, with dashed lines indicating the medians in each panel. In black, we show the selected comparison sample from the GAMA survey with S\'ersic profile fits in the $r$-band by \citet{Kelvin2012}. All distributions are normalised such that their integral is equal to 1. The light and stellar mass profiles of the EAGLE galaxies are skewed toward low S\'ersic indices, with only a slight increase in the median value of $n$ toward higher stellar mass. In comparison with the GAMA data, EAGLE is deficient in bulge-like ($n\sim4$) systems. The discrepancy between the distributions becomes stronger at higher stellar masses, and suggests a fundamental difference in the stellar mass density profiles of simulated and observed galaxies.}
    \label{fig:sersic_index}
\end{figure*}

The S\'ersic index characterises the shape of the surface brightness profile: a value of $n\sim1$ describes an exponential profile, often found in late-type galaxies, whereas local early-type galaxies tend to be well approximated by profiles with a value of $n\sim4$ (de Vaucouleurs profile). 
In Fig.~\ref{fig:sersic_index}, we show the probability distributions of the S\'ersic indices measured from the $r$-band (orange) and stellar mass (grey) images. Observationally, the shape of the surface brightness profile has been shown to correlate with different physical properties, such as the mass (or luminosity), colour and spectral age indicators \citep[e.g.,][]{Blanton2003,Kauffmann2003}. The sample is therefore split into three bins in stellar mass ($M_{\rm *,30}$), with dashed lines showing the median values in each panel. 

All mass bins show distributions that are skewed toward low values of $n$, indicating that the majority of galaxies are best described by profiles that closely resemble exponential discs. However, there is also an extended tail toward higher $n$, representing bulge-like profiles, which becomes more prominent at higher masses. This mass dependence is also apparent in the evolution of the median, as this increases from $n\approx 1.5$ ($n\approx1.8$) in the lowest mass bin to $n\approx2.0$ ($n\approx2.8$) in the highest mass bin for the $r$-band (stellar mass) fits.

Interestingly, the $r$-band imaging shows systematically different S\'ersic indices from the stellar mass imaging. The stellar mass profiles tend to be more concentrated in the centre, particularly at high stellar mass, with profiles that are closer to a classical de Vaucouleurs profile. In Appendix~\ref{sec:smoothing_lengths} we demonstrate that this is not due to the smoothing lengths used to create the $r$-band images, as we find identical results for a smoothed version of the stellar mass images. Rather, colour gradients appear to have a strong effect on the shape of the light profile, as was also noted by \citet{Kelvin2012}, who found systematic differences between their measurements of $n$ in the $r$-band and at near-infrared wavelengths (e.g., the $K$-band). Younger stellar populations at larger radii have low $M_*/L$, particularly at shorter wavelengths, which may drive the S\'ersic fit to lower observed values of $n$ than expected from the underlying stellar mass profile. This is in line with the findings by \citet{Trayford2019}, who showed that, based on the orbital properties of the stellar particles, younger stellar populations within EAGLE galaxies tend to reside in discs. Similarly, the effects of dust attenuation in the centre of the galaxy likely result in lower S\'ersic indices at rest-frame optical wavelengths.

To compare more directly with observational data, we use the catalogue of single S\'ersic profile fits to reprocessed SDSS $r$-band imaging from \citet{Kelvin2012}. We select all galaxies within $0.06<z<0.12$ and match this morphological catalogue with the stellar masses from \citet{Driver2016}, which were estimated using \textsc{MAGPHYS} \citep{daCunha2008}. The stellar masses are then scaled to the total flux of the best-fit S\'ersic profiles, and we select only galaxies with $\log(M_*/{\rm M_\odot})\geq 10$. We exclude galaxies with S\'ersic indices outside of $0.2<n<8$ for fair comparison with our own sample. Moreover, to filter out poor fits, we require that the reduced chi-squared value of the primary galaxy is within $0.5<\chi^2_{\nu, \rm pri}<2$. These criteria result in a final catalogue of 6554 GAMA galaxies with a median redshift of $z\approx0.1$.

The GAMA survey is highly complete for the selected mass and redshift range, and can therefore readily be compared with the EAGLE sample, which is by construction volume-limited. The distributions of the S\'ersic indices of the GAMA galaxies are shown in black in Fig.~\ref{fig:sersic_index}. For the lowest masses ($M_*\sim10^{10.2}\,\rm M_\odot$), there is reasonable agreement between the observed and simulated $r$-band data, as both distributions peak around $n\sim1.5$. However, the GAMA data show a less strongly peaked distribution at low $n$, and a more significant tail toward $n\sim4$. This large number of bulge-like galaxy profiles is simply missing in EAGLE, and this discrepancy becomes even stronger at higher masses, where GAMA consists predominantly of high $n$ systems.
The fact that these discrepancies remain when comparing the stellar mass values of $n$ with the GAMA data, suggests that it is the intrinsic mass distribution that differs from observations, rather than potential issues in the forward modelling (e.g., uncertainties in the dust properties and geometry). We further discuss the discrepant mass distribution of simulated galaxies in Section~\ref{sec:discussion_profiles}.

\subsection{Axis ratios}\label{sec:axis_ratio}

\begin{figure*}
    \centering
    \includegraphics[width=\linewidth]{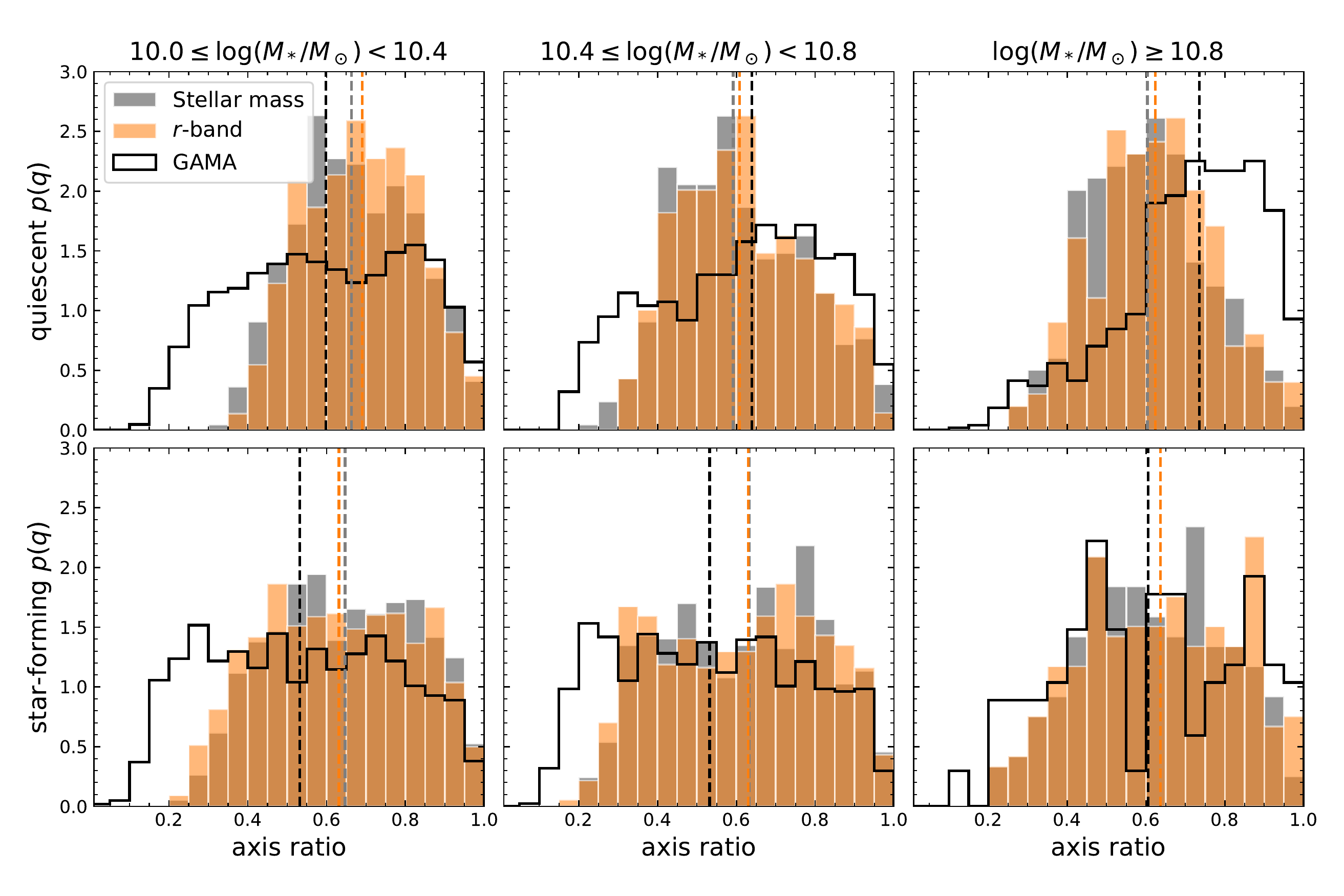}
    \caption{Projected axis ratio distributions of the EAGLE and GAMA comparison galaxies in bins of increasing stellar mass, further separated into quiescent (top panels) and star-forming (bottom panels) subsamples. Symbols indicate the same as in Fig.~\ref{fig:sersic_index}. Unlike in Fig.~\ref{fig:sersic_index}, the stellar mass and $r$-band fits show good agreement in all mass bins, and for both star-forming and quiescent galaxies. In line with observations, the quiescent subsamples show more strongly peaked distributions than the star-forming subsamples, which is consistent with a higher proportion of disc-like intrinsic shapes among the star-forming population. The main difference between the simulated and GAMA data is at low axis ratios, as there are no highly flattened systems in EAGLE, likely due to the imposed gas pressure floor and the limited resolution of the simulation. In the highest mass bin, the quiescent GAMA sample is skewed toward higher axis ratios, which implies that the GAMA galaxies are intrinsically rounder in shape than the EAGLE galaxies. }
    \label{fig:axis_ratio}
\end{figure*}

The second morphological parameter is the ratio between the semi-major and semi-minor axes, which provides insight into the intrinsic shape of a system. However, due to projection effects, this cannot be done on an object-by-object basis. Rather, it is the distribution of axis ratios that is often used to infer the distribution of the intrinsic shapes for a sample of galaxies \citep[see, e.g.,][]{Holden2012,Chang2013,vdWel2014b}.

As this is typically done separately for star-forming and quiescent galaxy populations, we divide our sample by the sSFR, as in Section~\ref{sec:mass_size}. For GAMA we use the sSFR averaged over the last 100\,Myr from the \textsc{MAGPHYS} SED modelling. The projected axis ratio distributions for these two populations are shown in Fig.~\ref{fig:axis_ratio}, in bins of stellar mass. The orange and grey histograms show the $r$-band and stellar mass results for EAGLE, respectively, and black corresponds to the GAMA results; dashed lines indicate median values.

Using a Kolmogorov-Smirnov (KS) test, we find that both the $r$-band and stellar mass distributions agree very well. In contrast with the S\'ersic index distributions (Fig.~\ref{fig:sersic_index}), which show that the stellar mass and light density profiles differ significantly, the stellar mass and light do trace the overall (3D) shapes of galaxies in the same manner. Moreover, we find that the star-forming and quiescent EAGLE galaxies indeed follow significantly ($>3\sigma$) different distributions, apart from the highest mass bin, where the number of galaxies is also substantially smaller ($\sim 200$ versus $\sim700$ at lower masses). 
The distribution of quiescent EAGLE galaxies shows a peak around $q\approx0.65$ in each panel, whereas the star-forming galaxies show a more uniform spread, which may be explained by a larger proportion of disc-like (oblate) systems within the star-forming population. 

On the other hand, the median values of the distributions do not differ strongly between the two populations, nor show a dependence on the stellar mass, both of which are clear features in the GAMA data. The most significant difference between the observed and simulated data is at low axis ratios: a large number of GAMA galaxies are highly flattened (with $q\approx0.2$), yet, these galaxies do not exist in the EAGLE simulations. As discussed by \citet{Trayford2017} and \citet{vdSande2019}, galaxies in EAGLE tend to be thicker than is observed, possibly as the result of the pressure floor that is imposed within the simulation. Moreover, the limited mass resolution of the dark matter particles in the simulation has been shown to lead to a heating of the baryonic particles via 2-body scattering \citep{Ludlow2019,Ludlow2021}.   

Furthermore, there is a discrepancy between the GAMA and EAGLE data at high $q$ for the quiescent galaxies. Whereas the EAGLE data show very little dependence on mass, the axis ratio distribution in GAMA is increasingly skewed toward high $q$ at higher masses. This difference is most apparent in the highest mass bin, which peaks at $q\approx0.8$ for the GAMA galaxies. \citet{Chang2013} showed that low-redshift galaxies in this mass range are mainly triaxial systems ($\approx 80\%$ of the sample) with a mean intrinsic major-to-minor axis ratio of $C/A\approx0.6$ and mean triaxiality parameter $T\approx0.6$. For EAGLE, the highest mass bin may still contain a significant number of triaxial systems, but with more flattening along the intermediate or minor axes. Indeed, based on the 3D stellar mass distribution, \citet{Thob2019} showed that there is a significant population of triaxial systems and prolate systems in EAGLE, with $C/A\sim0.4$ among galaxies in the red sequence. Contrary to the intermediate mass galaxies ($\log(M_*/{\rm M_\odot})\lesssim 10.8$) that are not flattened enough, at the highest masses the simulation thus struggles to reproduce galaxies that are sufficiently round.

\section{Discussion}\label{sec:discussion}

\subsection{The importance of apples-to-apples comparisons}

The differences and similarities found between the structural properties of simulated and observed galaxies, only truly become apparent when using mock observations and the same measurement techniques as in large galaxy surveys. This can also lead to different conclusions: \citet{Correa2017} showed, using the stellar kinematics to identify spheroids and discs, that kinematic morphologies in EAGLE are tightly correlated with the $u-r$ colour, with central galaxies along the red sequence being dominated by spheroidal morphologies. Although we find some dependence of $q$ and $n$ on stellar mass and sSFR, the lack of $n\sim4$ systems would lead to a different picture of the red sequence. 

Perhaps more important, however, is the remarkable improvement in the stellar mass-size relation when $M_*/L$ gradients are modelled, and the definition of galaxy size is made consistent with observations. The mass-size relation is often used as a key measure of success for cosmological simulations, and in the case of EAGLE also plays a role in the calibration of the subgrid model.

Yet, we have found that the 3D curve of growth methods commonly used to measure half-mass or half-light radii differ systematically from the semi-major axis sizes obtained with 2D S\'ersic modelling. It is therefore difficult to directly compare the resulting mass-size relation with observations. Comparison with circularised sizes ($r_{\rm e,circ}\equiv\sqrt{q}\, r_{\rm e,maj}$), as done by, e.g., \citet{Genel2018} or \citet{vdSande2019} using 2D growth-curve sizes, is possibly even more complex, as there is an additional dependence on the distribution of the projected axis ratios. Rather, the semi-major axis is the preferred measure of size here, as it is largely independent of inclination and intrinsic axis ratio for oblate systems, which is the most commonly found shape of $z\sim0$ galaxies \citep[][ although the effects of dust complicate this slightly, as demonstrated in Fig.~\ref{fig:delta_re}]{Chang2013,vdWel2014b}. When comparing with observations, this therefore allows to distinguish between a possible systematic offset in the sizes and a mismatch in the distribution of the intrinsic shapes. 

By making a consistent comparison using the semi-major axis sizes, we have shown that the half-mass radii turn out systematically smaller than measurements in the $r$-band from GAMA \citep{Lange2015}, whereas there is significantly better agreement with observations for the quiescent population. 
The excellent agreement found previously between the 3D half-mass radii and observed $r$-band sizes of star-forming galaxies \citep{Furlong2017} is therefore partially the result of the model calibration, and to some extent simply coincidence. 

However, accounting for gradients in $M_*/L$ with the use of the mock $r$-band imaging brings the star-forming population in good agreement again with the observed relation, and with a scatter that is closer to that observed. As also shown by \citet{vdSande2019}, the effect of using luminosity-weighted sizes rather than mass-weighted sizes is significant, and is further enhanced by the implementation of realistic dust attenuation in this work \citep[see also][ for the effects of dust on the measurement of structural parameters]{Gadotti2010}. On the other hand, the location of the quiescent population is changed only minimally within the mass-size plane, consistent with the expectation that these galaxies have less variation in $M_*/L$. 

Remaining discrepancies, the sizes of both star-forming and quiescent EAGLE galaxies are approximately 0.1\,dex larger at fixed mass, can be caused by a large number of factors within the simulation itself. Additionally, uncertainties in the radiative transfer modelling (e.g., the treatment of molecular clouds in the ISM) may introduce a systematic uncertainty on the $M_*/L$ gradients, and hence the size measurements. The deviating shapes and morphologies of the simulated galaxies will also affect the simulated $M_*/L$ gradients, as the results of the radiative transfer calculations are dependent on the geometry of both the stellar particles and dust \citep[e.g., for the difference in the dust attenuation between thin and thick discs, see][]{Trayford2017}.
Finally, it is also important to bear in mind that the stellar masses inferred with SED modelling carry large uncertainties \citep[$\approx 0.3\,$dex at $z\sim0$;][]{Conroy2009}, which can introduce a systematic uncertainty of similar magnitude in the observed mass-size relation \citep[see][]{Genel2018}.

\subsection{Mismatched density profiles and intrinsic shapes}\label{sec:discussion_profiles}

We have demonstrated that the morphologies of EAGLE galaxies, as quantified by the S\'ersic index, differ significantly from observations: at all stellar masses ($\log(M_*/{\rm M_\odot})>10$), there are too few galaxies with bulge-like ($n\sim4$) light profiles. The fact that this discrepancy holds true also for the stellar mass surface density profiles, shows that the mass is distributed differently in simulated galaxies, and that observational effects (measuring light versus stellar mass, effects of dust attenuation) are of secondary importance.

\begin{figure}
    \centering
    \includegraphics[width=\linewidth]{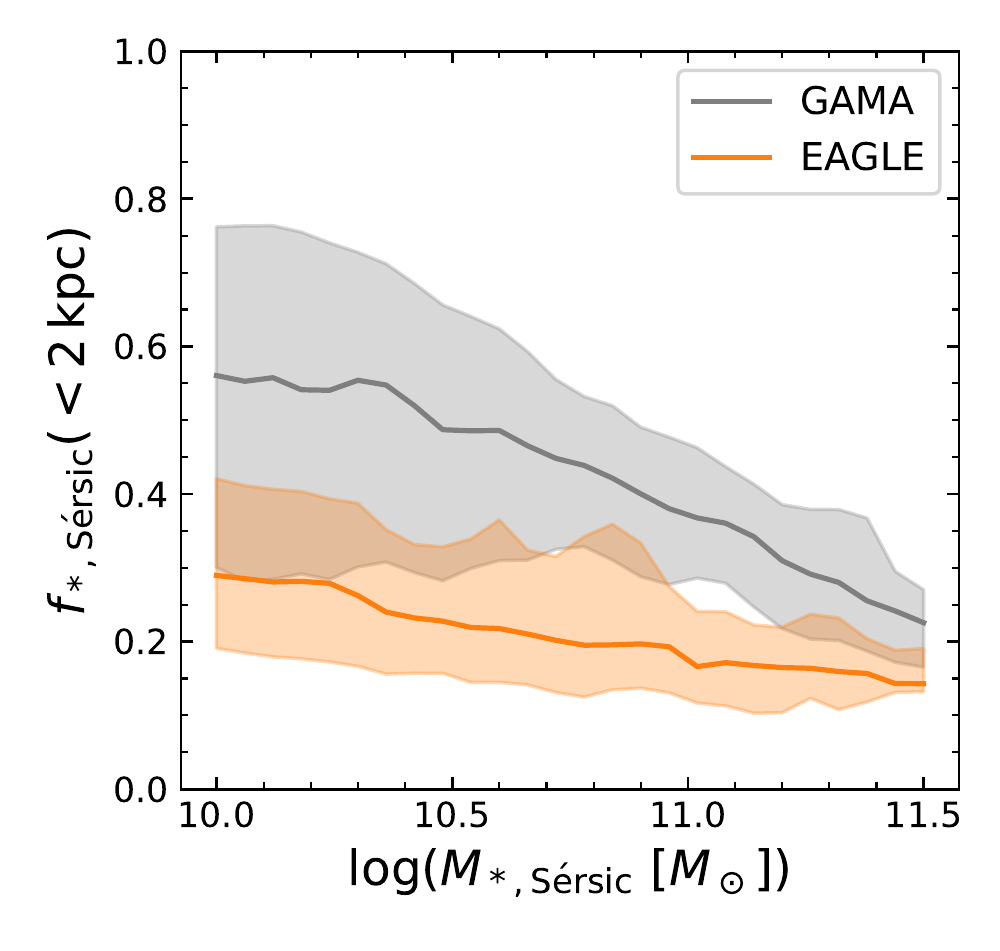}
    \caption{The stellar mass fraction within an aperture of radius 2\,kpc measured from the best-fit S\'ersic model, as a function of the total stellar mass inferred from the same model. Solid lines show the running median of the EAGLE (orange) and GAMA (grey) data, and shaded regions mark the 16th to 84th percentile range. In comparison with observations, the inferred stellar mass density profiles of EAGLE galaxies are deficient in mass at small scales.}
    \label{fig:flux_deficit}
\end{figure}

We highlight this finding in Fig.~\ref{fig:flux_deficit}, where we show the fraction of stellar mass enclosed within a fixed aperture of radius 2\,kpc as a function of the total stellar mass of the best-fit model. Both the mass fractions and total masses are inferred from the best-fit S\'ersic models, therefore demonstrating the physical difference between the S\'ersic index distributions in EAGLE and GAMA: the stellar mass fractions in EAGLE are a factor $\approx2$ below the observed mass fractions in GAMA, irrespective of the stellar mass.

A deficiency in bulge-like systems was also found by \citet{Bottrell2017a,Bottrell2017b}, who constructed mock SDSS observations of galaxies in the Illustris simulations \citep{Vogelsberger2014} and performed a two component (bulge + disc) S\'ersic profile fitting to determine the bulge fractions.
As also noted by \citet{Bottrell2017b}, contrary to issues with early hydrodynamical simulations producing galaxies that were too bulge-like \citep[e.g.,][]{Katz1991}, it therefore appears that some of the more recent models struggle to form enough bulges. \citet{Rodriguez2019} showed, using single S\'ersic profile fits to mock Pan-STARRS imaging, that this is also the case for simulated galaxies in Illustris-TNG: although the galaxy sizes are in significantly better agreement with observations than was the case for Illustris, the morphologies are still too disc-like, as galaxies below a mass of $\log(M_*/{\rm M_\odot})\approx 10.7$ follow profiles with $n\approx 1.5$. Only for the most massive galaxies ($\log(M_*/{\rm M_\odot})\gtrsim 11$) is there good agreement between the simulated and observed S\'ersic indices. 

Interestingly, non-parametric methods of quantifying morphology paint a slightly different picture. Based on the same optical imaging used in this work, but with slightly different noise and instrument resolution applied, \citet{Bignone2020} find that the distributions of nearly all commonly used non-parametric measures \citep[e.g., Gini coefficient, Concentration parameter;][]{Lotz2004} match well with observations from the GAMA survey. The non-parametric morphologies of galaxies in Illustris-TNG \citep{Rodriguez2019} are approximately equally successful in reproducing observations, demonstrating a substantial improvement with respect to previous measurements from the Illustris simulations \citep{Snyder2015}.

Based on the discrepant S\'ersic indices alone, it may be tempting to conclude that there are improvements to be made in the physics implemented in the simulations, such as the feedback prescriptions. Yet, the non-parametric morphologies do not show a strong indication for this, particularly in EAGLE, where the non-parametric measures additionally correlate with the stellar mass and sSFR in the same way as in observations.

Reconciling the different outcomes of these two strategies of measuring galaxy morphologies is not immediately obvious, however, the residual surface brightness and density profiles may provide some insight. As shown in Fig.~\ref{fig:sb_profs}, there are minor, systematic features in the residual flux (i.e., the difference between the mock image and best-fit S\'ersic model). 
Although the excess low surface brightness emission at large radii ($\gtrsim 3\,r_{\rm e}$) could simply reflect the fact that a two-component model (bulge+disc decomposition) would be a better description for the galaxy profiles, the features at smaller radii are not as easily `fixed'.

The under-subtraction at $r<r_{\rm e}$ ($\approx 2\,$kpc) followed by over-subtraction at $r\approx 1-2\, r_{\rm e}$ ($\approx 5\,$kpc) suggests that the surface brightness profile declines more steeply than a $n\sim2$ model describes. Similarly, the excess flux at $r\gtrsim 3\,r_{\rm e}$ indicates that the profile is shallower than a $n\sim2$ profile at large radii. A steep decline at small radii followed by a gradual decline at large radii is characteristic of a high S\'ersic index profile ($n\gtrsim4$). 
However, likely due to the high S/N in the central pixels, the fit is driven to low S\'ersic indices. It therefore appears that the simulated galaxies are simply deficient in mass and light at the very centre in comparison to the rest of the galaxy, which may be the effect of the resolution limit in the simulation \citep[see also][]{Schaller2015} and the associated 2-body scattering of dark matter and baryonic particles \citep{Ludlow2019,Ludlow2021}. The pressure floor within the simulation may also play a role here, as the associated spatial scale of $\sim1\,$kpc likely affects the inner density mass density profiles, and therefore the measured S\'ersic indices. These effects may also explain the similarities between the sizes and morphologies in the EAGLE and Illustris-TNG simulations \citep{Genel2018,Rodriguez2019}, as, although the two simulations employ different physical models (e.g., the feedback prescriptions), both use a similar resolution and pressure floor. 

A bulge+disc decomposition then also does not offer substantial improvement, because the profile shape in the centre deviates too strongly from a S\'ersic profile. On the other hand, within the apertures used to calculate non-parametric morphological measures, these features in the light profiles may be washed out, and thus provide an explanation for their better consistency with observations. Restricting the S\'ersic profile fitting to $r>2\,$kpc may then be an appropriate method to minimise the effects of the unrealistic profile shapes in the centres.

This breakdown in the density profiles at small scales would also help to reconcile the discrepant results found between the S\'ersic index distributions and the projected axis ratios, as we may expect galaxies with $n\sim1-2$ profiles to have disc-like (oblate) intrinsic shapes. Whereas this is likely the case for the star-forming population (Fig.~\ref{fig:axis_ratio}), which only differ from observations by the lack of highly flattened galaxies, the quiescent galaxies show projected axis ratios that more plausibly reflect a large population of triaxial and prolate systems, as also shown to be present in EAGLE by \citet{Thob2019} and \citet{Trayford2019}.

Although the quiescent galaxies are not as round as seen in observations, particularly at high stellar mass, the projected axis ratios show a picture that is closer to reality than would be concluded from the S\'ersic indices alone, and is more consistent with the variety of bulges and discs found in studies that use kinematic morphologies as a proxy for the observed morphology \citep[e.g.,][]{Correa2017,Clauwens2018}. This then raises the question of whether S\'ersic indices have any predictive power for the intrinsic shapes of simulated galaxies. 

\begin{figure}
    \centering
    \includegraphics[width=\linewidth]{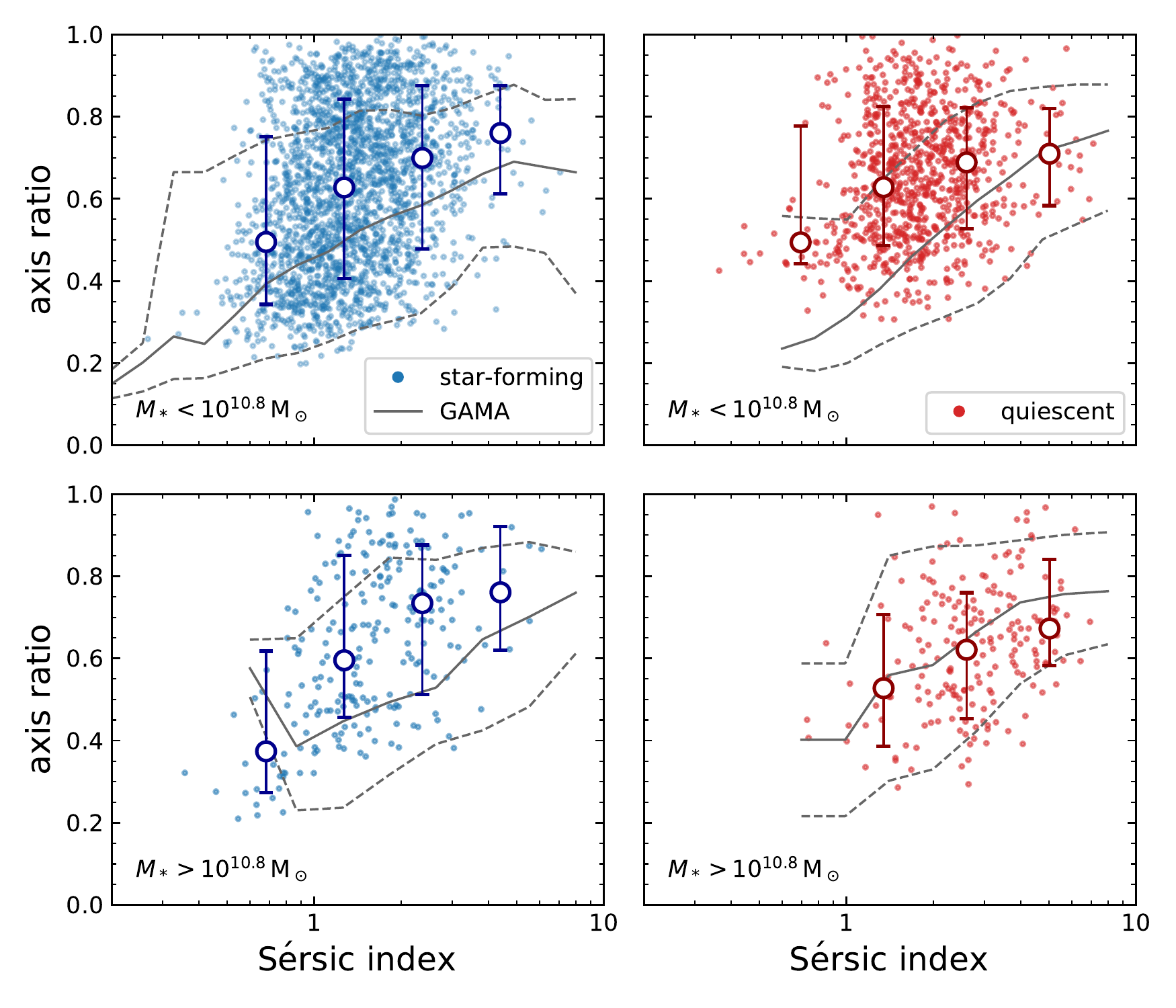}

    \caption{The $r$-band projected axis ratio versus S\'ersic index, for star-forming (blue) and quiescent (red) EAGLE galaxies in two different stellar mass ranges (intermediate masses in top panels; high masses in bottom panels). Open circles show the median axis ratios of the EAGLE galaxies in logarithmic bins in S\'ersic index, with error bars indicating the 16th and 84th percentiles. Grey lines show the 16th, 50th and 84th percentiles for the GAMA survey. The axis ratio generally increases toward higher S\'ersic index, consistent with the picture of $n\sim4$ galaxies being rounder in shape, although the scatter is large. Whereas Fig.~\ref{fig:axis_ratio} showed that massive quiescent galaxies in GAMA are intrinsically more round than is the case in EAGLE, this is partly due to the difference in the S\'ersic index distributions: the shapes of the few bulge-like quiescent galaxies in EAGLE are in good agreement with observations.}
    \label{fig:axis_ratios_bulges}
\end{figure}

In Fig.~\ref{fig:axis_ratios_bulges} we show the projected axis ratios as a function of the S\'ersic index for the $r$-band S\'ersic models, with different panels separating the star-forming (left) and quiescent (right) populations, as well as massive and less-massive galaxies (top versus bottom panels, the boundary used being $M_*=10^{10.8}\,\rm M_\odot$). In addition to showing the individual EAGLE galaxies, the open circles show the median axis ratios in bins of S\'ersic index (with error bars showing the 16th and 84th percentiles). For comparison, we also show the running median for the comparison sample from GAMA (solid lines), with dashed lines indicating the 16th and 84th percentiles. 

For the star-forming galaxies, the EAGLE data show a positive correlation between axis ratio and S\'ersic index, albeit with large scatter. This is similar to the correlation and scatter present in the GAMA data, except for an offset toward slightly higher axis ratio at fixed S\'ersic index, which is likely explained by the lack of thin discs in EAGLE (as also discussed in Section~\ref{sec:axis_ratio}). 

The quiescent galaxies of $M_*<10^{10.8}\,\rm M_\odot$ (top right panel) do not show such a clear correlation, and thereby diverge from the trend seen in GAMA. On the other hand, the massive quiescent galaxies (bottom right panel) do show a slight increase in the median axis ratio toward higher S\'ersic indices, and follow the observed correlation almost exactly, although the number of EAGLE galaxies (199) in this panel is relatively small. 

The EAGLE simulations thus do appear to produce galaxies that resemble the classical picture of spheroidal galaxies with $n\sim4$ light profiles, only not in sufficient number. As suggested previously, the resolution or the pressure floor may play a role at the small scales probed with these density profiles. Other, more physical effects could be in the implementation of the central black hole and stellar feedback in the simulations: the orbital structure of both the dark matter and stellar particles depend on the feedback mechanisms employed, with strong (black hole) feedback resulting in a higher fraction of box orbits and thus more strongly triaxial systems \citep{Bryan2012}. The observed axis ratio distribution may therefore offer an interesting constraint on the subgrid model, being an observable that is not as model-dependent as the light profile shape or definition of size.

\section{Conclusions}\label{sec:conclusion}

Starting from the optical images of $z=0.1$ EAGLE galaxies constructed with \textsc{skirt} by \citet{Trayford2017}, we have created mock $r$-band images that have similar noise properties and resolution as photometric data from the SDSS. Following methods that are commonly used in observational studies, we have fitted S\'ersic profiles to these mock observations using a combination of \textsc{SExtractor} and \textsc{Galfit}, thus enabling an apples-to-apples comparison between the structural parameters of galaxies in EAGLE and local observations.

To be able to distinguish between the effects of different measurement techniques and the effects of variations in $M_*/L$ (due to, e.g., recent star formation or dust attenuation), we have constructed a second set of images from the projection of the stellar mass particles. These stellar mass images are created such that the noise and resolution match the mock optical images.

Our findings can be summarised as follows:
\begin{itemize}
    \item Galaxy sizes depend on the measure of size used, as there are systematic differences between the half-mass radii estimated with a curve-of-growth method (common in theoretical work) and the semi-major axes obtained with S\'ersic profile modelling (common in observational studies). The magnitude of this discrepancy is on average $\approx0.06\,$dex, but is itself dependent on the galaxy size.
    \item Gradients in $M_*/L$ due to radial variations in the star formation, stellar age and dust attenuation can have a large effect on the observed size: half-light radii are typically 25\% larger than half-mass radii, but with large scatter and outliers that deviate by as much as a factor $\approx3$. For quiescent galaxies, on the other hand, the light-weighted structural properties provide a good proxy of the mass-weighted properties.
    \item The measured stellar mass-size relation thus also depends strongly on the method used to determine the size (and corresponding stellar mass). Only for the $r$-band half-light radii estimated with the S\'ersic modelling, is the mass-size relation in EAGLE in good agreement with observations for both star-forming and quiescent galaxies, albeit with a systematic offset of 0.1\,dex.
    \item The S\'ersic indices of EAGLE galaxies tend be lower than observed, due to a deficiency in bulge-like ($n\sim4$) systems. A closer look at the surface brightness and mass density profiles shows that there is likely a deficiency in stellar mass (and hence light) at the very centres of the simulated galaxies.
    \item There is a lack of highly flattened objects among both the quiescent and star-forming population, likely due to the gas pressure floor and the limited resolution of the simulation. On the other hand, massive quiescent galaxies in EAGLE are not sufficiently round in shape, and appear to be more strongly triaxial than quiescent galaxies in GAMA.

\end{itemize}

Our work demonstrates that, for a fair comparison between the structural parameters of simulated and observed galaxies, it is crucial to account for the effects of $M_*/L$ gradients within galaxies, as well as the systematic differences between various analysis techniques. This can be achieved by either deriving mass-weighted measurements from observations or, as shown here, by constructing realistic mock observations from simulations.
Although computationally expensive, a realistic treatment of simulated data can truly provide a different picture of the simulated galaxy population.

\section*{Acknowledgements}

AdG thanks Sarah Appleby for useful discussions and her help in recovering an early version of the modelling software. 
MS is supported by the Netherlands Organisation for Scientific Research (NWO) through VENI grant 639.041.749. 

We acknowledge the Virgo Consortium for making their simulation data available. The EAGLE simulations were performed using the DiRAC-2 facility at Durham, managed by the ICC, and the PRACE facility Curie based in France at TGCC, CEA, Bruy\`eres-le-Ch\^atel.

GAMA is a joint European-Australasian project based around a spectroscopic campaign using the Anglo-Australian Telescope. The GAMA input catalogue is based on data taken from the Sloan Digital Sky Survey and the UKIRT Infrared Deep Sky Survey. Complementary imaging of the GAMA regions is being obtained by a number of independent survey programmes including GALEX MIS, VST KiDS, VISTA VIKING, WISE, Herschel-ATLAS, GMRT and ASKAP providing UV to radio coverage. GAMA is funded by the STFC (UK), the ARC (Australia), the AAO, and the participating institutions. The GAMA website is http://www.gama-survey.org/ .





\bibliographystyle{mnras}
\bibliography{sims} 



\appendix

\section{Impact of particle smoothing}\label{sec:smoothing_lengths}

\begin{figure*}
    \centering
    \includegraphics[width=\linewidth]{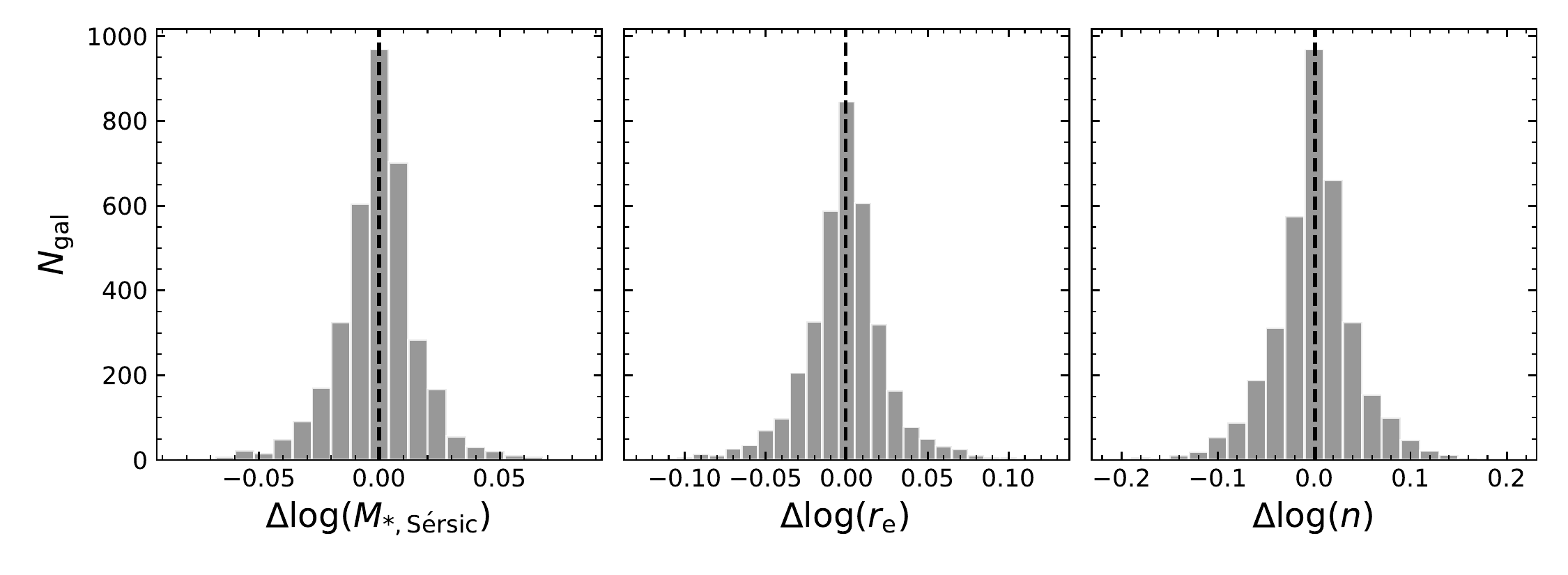}
    \caption{Differences between the structural parameters measured from stellar mass images with and without nearest neighbour smoothing. Median values are indicated with dashed lines. There is no systematic offset in the inferred stellar mass, half-mass radius and S\'ersic index, and the scatter is consistent with the uncertainties discussed in Section~\ref{sec:galfit_errors}. The differences between the $r$-band and stellar mass sizes and morphologies presented in this paper are therefore unlikely to be caused by a difference in the smoothing of the stellar particles.}
    \label{fig:smoothing}
\end{figure*}

As described in Section~\ref{sec:skirt}, the images created with \textsc{skirt} assume a truncated Gaussian profile for the spatial distribution of the stellar particles. The width of this distribution, the smoothing length, is set equal to the distance to the 64th nearest neighbour particle. On the other hand, to create the stellar mass images (Section~\ref{sec:mock_im_mstar}), the stellar particles were treated as point sources.

If the effect of smoothing is large, we may expect to find less centrally concentrated light profiles, and thus lower S\'ersic indices. As the size and total luminosity covary with the S\'ersic index, this may also affect the obtained mass-size relation.

Given the computational cost of creating mock $r$-band images, we evaluate the effect of different smoothing lengths on the stellar mass images instead. We use the \textsc{py-sphviewer} code \citep{sphviewer} to construct stellar mass maps that include nearest neighbour smoothing. These images are created from the exact same particles used before, and thus have identical dimensions and orientation. The code then uses the 3D particle distribution to compute the distance to the 64th nearest neighbour for all particles, and provides a smoothed, projected image of $512\times512$ pixels. Finally, we process this image in the same fashion as described in Section~\ref{sec:mock_im_mstar} to apply PSF smoothing, degrade the pixel resolution, and add realistic noise.

We perform the S\'ersic profile modelling for this set of smoothed images, and compare the resulting structural parameters to the fits without smoothing. Fig.~\ref{fig:smoothing} shows the distributions of the differences in the obtained stellar masses, half-mass radii, and S\'ersic indices. All three distributions are centred around zero (median values indicated with dashed lines), and the scatter is consistent with the typical uncertainties discussed in Section~\ref{sec:galfit_errors}. We can therefore conclude that the differences found in the structural parameters measured from the $r$-band and stellar mass images are not due to a difference in the applied smoothing.

\section{Functional form of the PSF}\label{sec:apdx_psf}

To create realistic mock observations, we convolved the images with a PSF that takes the form of a circular Gaussian distribution (Section~\ref{sec:mock_sdss}):
\begin{equation}
    {\rm PSF}(\mathbf{r}) =  {\rm PSF}(r) = \frac{1}{2\pi\sigma^2}  \exp{ \left (  \frac{- r^2}{2\sigma^2} \right)}\,, 
\end{equation}
where the variance $\sigma^2$ is related to the width of the distribution by ${\rm FWHM} = 2\sqrt{2\ln{2}}\times \sigma$. 

However, the shape of the PSF is generally more complex than a single Gaussian distribution describes, as, in addition to a core component, there are typically extended wings. In the SDSS, the PSF has been modelled with various decompositions \citep[see][]{Stoughton2002}, the simplest being a double Gaussian model, which was also used to perform surface brightness profile fitting to postage-stamp images of galaxies (to obtain `model magnitudes'). This double Gaussian model is isotropic and takes three parameters:
\begin{equation}
    {\rm PSF}(r) = \frac{(1-C)}{2\pi\sigma_1^2}  \exp{ \left (  \frac{- r^2}{2\sigma_1^2} \right)} + \frac{C}{2\pi\sigma_2^2}  \exp{ \left (  \frac{- r^2}{2\sigma_2^2} \right)}\,, 
\end{equation}
where $\sigma_1^2$ and $\sigma_2^2$ are the variances of the two components, and the constant $C$ is the ratio of the overall amplitudes. 

\begin{figure}
    \centering
    \includegraphics[width=\linewidth]{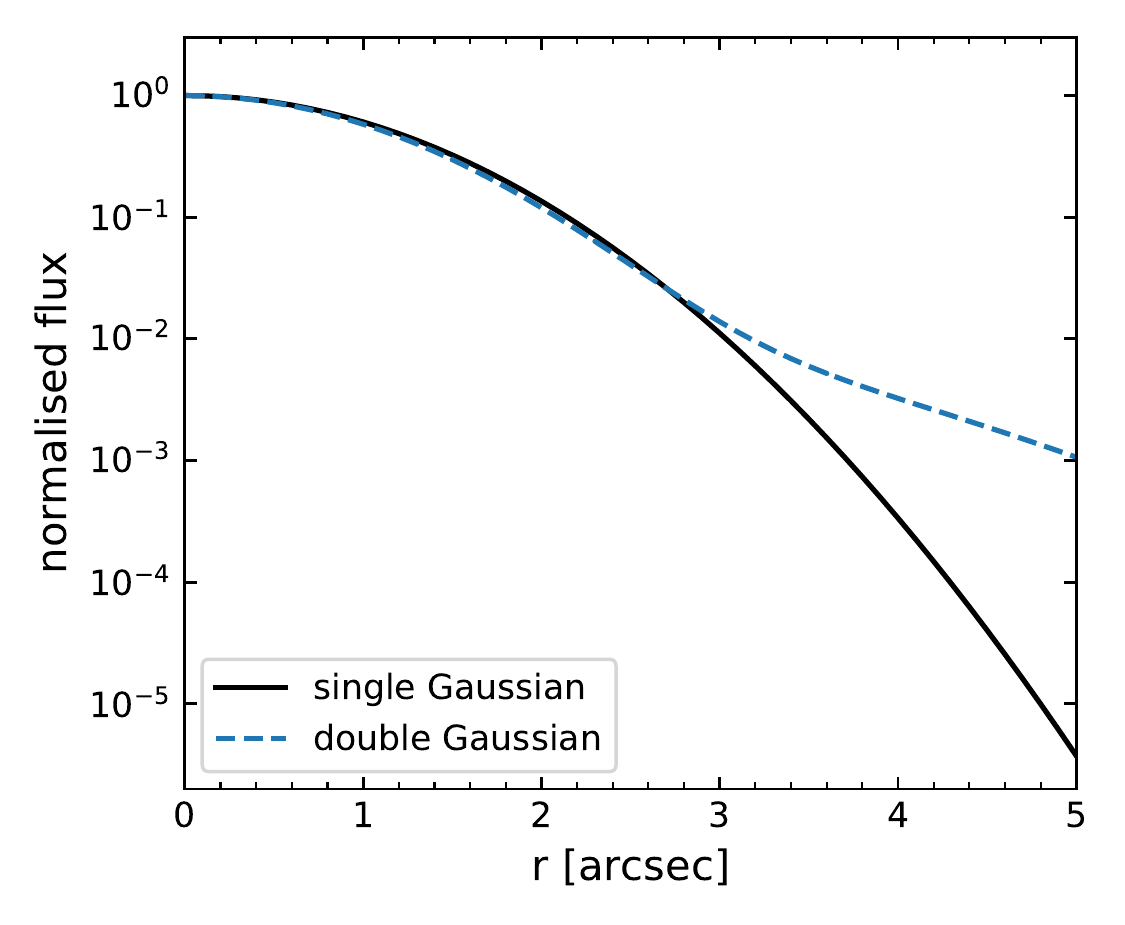}
    \caption{Radial profiles of the single Gaussian and double Gaussian models of the SDSS PSF. The two models are comparable at small scales, but deviate strongly in the outer wings.}
    \label{fig:psfs}
\end{figure}

\begin{figure*}
    \centering
    \includegraphics[width=0.92\linewidth]{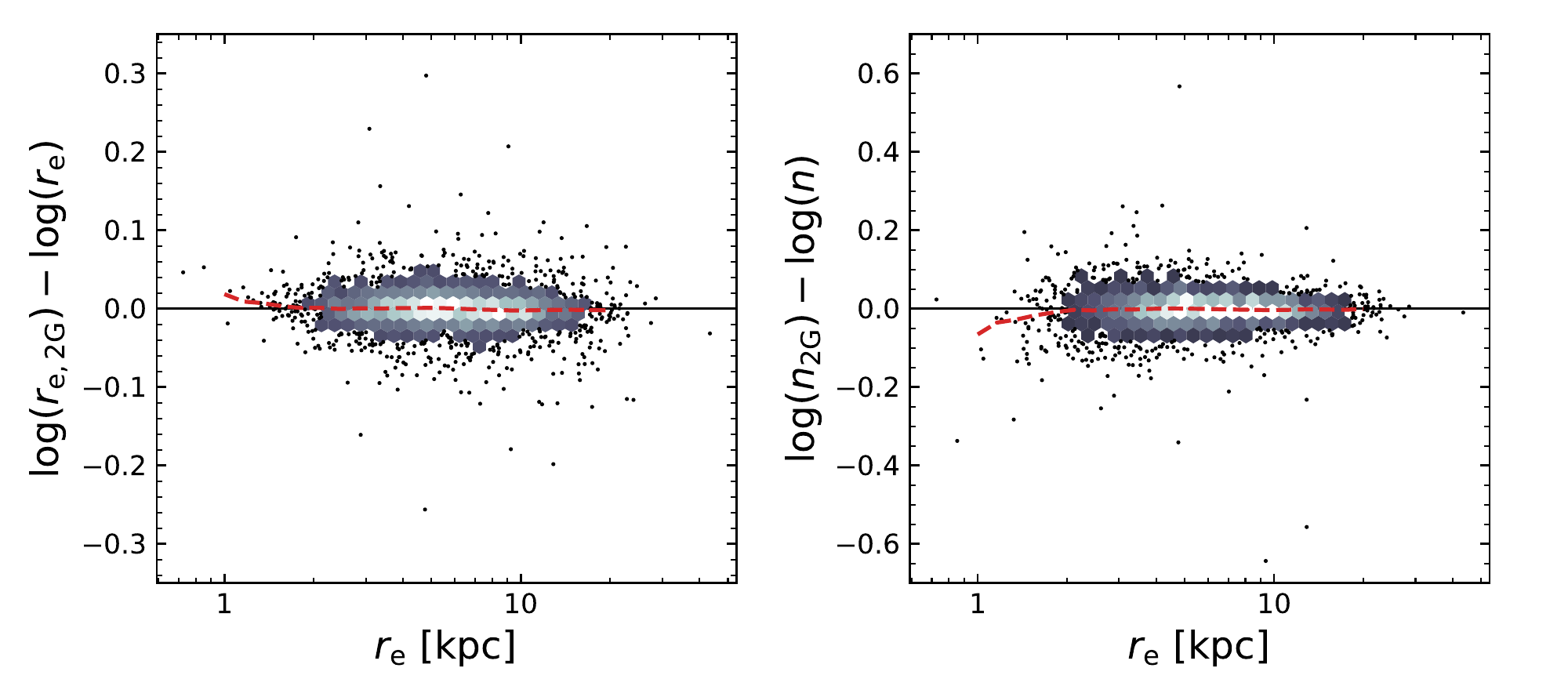}
    \caption{The difference in the inferred $r$-band size (left) and S\'ersic index (right) between the images that are convolved with a single Gaussian PSF and a double Gaussian PSF. There is no systematic offset between the two sets of models (medians are indicated with a red, dashed line), and the scatter is consistent with the random uncertainties discussed in Section~\ref{sec:galfit_errors}. Only for very small galaxies ($r_{\rm e}\lesssim2\,$kpc), where we may expect to see the largest impact of a difference in the PSF, does there appear to be a slight offset in both the size and S\'ersic index, of up to $\Delta\log(r_{\rm e})\sim0.05\,$dex and $\Delta\log(n)\sim -0.1\,$dex. The choice of adopting a simple (single Gaussian) PSF shape thus has minimal impact on the results presented in this work. }
    \label{fig:psf_galfit}
\end{figure*}

In Fig.~\ref{fig:psfs}, we show the radial profiles of these two different PSF shapes, using $\rm FWHM=1.39\arcsec$ for the single Gaussian PSF (solid black line). For the double Gaussian model (blue dashed line), we obtain typical parameters from the `Field' catalogue described in Section~\ref{sec:mock_ugriz}: $[\sigma_1,\sigma_2, C] \approx [0.95\sigma,2.05\sigma,0.09]$. There is good agreement between the two PSFs at small radii, with the FWHM of the double Gaussian model being only slightly smaller ($\rm FWHM = 1.33\arcsec$). For $r\gtrsim 3\arcsec$ the two models deviate increasingly strongly, by multiple orders of magnitude.

We quantify the difference between these two profile shapes by calculating the second moment of the distributions \citep[see also][]{Franx1989}:
\begin{equation}
    \langle r^2 \rangle = \frac{ \iint_{\mathbb{R}^2} r^2\, {\rm PSF}(\mathbf{r}) \,{\rm d}\mathbf{r} } { \iint_{\mathbb{R}^2} {\rm PSF}(\mathbf{r}) \,{\rm d}\mathbf{r} },
\end{equation}
which for the normalised, isotropic PSFs considered here reduces to
\begin{equation}
    \langle r^2 \rangle = 2\pi \int_0^{\infty} r^2\, {\rm PSF}(r) \,r\,{\rm d}r\,.
\end{equation}
For the single and double Gaussian models described previously, these moments are $\langle r^2 \rangle = 0.70\,$arcsec$^2$ and $\langle r^2 \rangle = 0.84\,$arcsec$^2$, respectively.

Whereas the difference in the FWHM is only 5\%, the second moments differ by 20\%, which may have a measurable effect on the constructed images and inferred structural parameters. This effect is expected to be largest for galaxies that have clumpy surface brightness profiles, as these would appear more smooth with the double Gaussian PSF, as well as for highly compact galaxies that would appear more extended. 

We therefore use the optical images (Section~\ref{sec:skirt}) to evaluate whether the choice of the PSF model leads to systematic effects. A second set of mock $r$-band images is constructed in the exact same way as described in Section~\ref{sec:mock_ugriz}, except for the use of the double Gaussian model in the PSF smoothing instead of the single Gaussian model. We then run the S\'ersic modelling pipeline on these images, and compute the difference in the obtained structural parameters. 

We focus on the size and S\'ersic index, as these are the parameters that are most likely to be affected by a change in the PSF. Fig.~\ref{fig:psf_galfit} shows the difference in the obtained half-light radius and S\'ersic index, as a function of the half-light radius. Generally, the two sets of measurements agree very well, as there is no systematic offset and little scatter (consistent with the expected measurement uncertainties). Only at very small radii ($r_{\rm e}\lesssim2\,$kpc) is there are slight difference between the two PSF models, as the use of the double Gaussian PSF leads to slightly larger sizes (up to $\Delta\log(r_{\rm e})\sim0.05\,$dex) and slightly lower S\'ersic indices (up to $\Delta\log(n)\sim-0.1\,$dex). 

Overall, we can therefore conclude that the shape of the PSF has a minimal influence on the inferred structural parameters, and that a simple PSF model is sufficient for measuring parametric morphologies.

\section{Comparing S\'ersic model stellar masses with aperture measurements}\label{apdx:aperture}

\begin{figure*}
    \centering
    \includegraphics[width=0.92\linewidth]{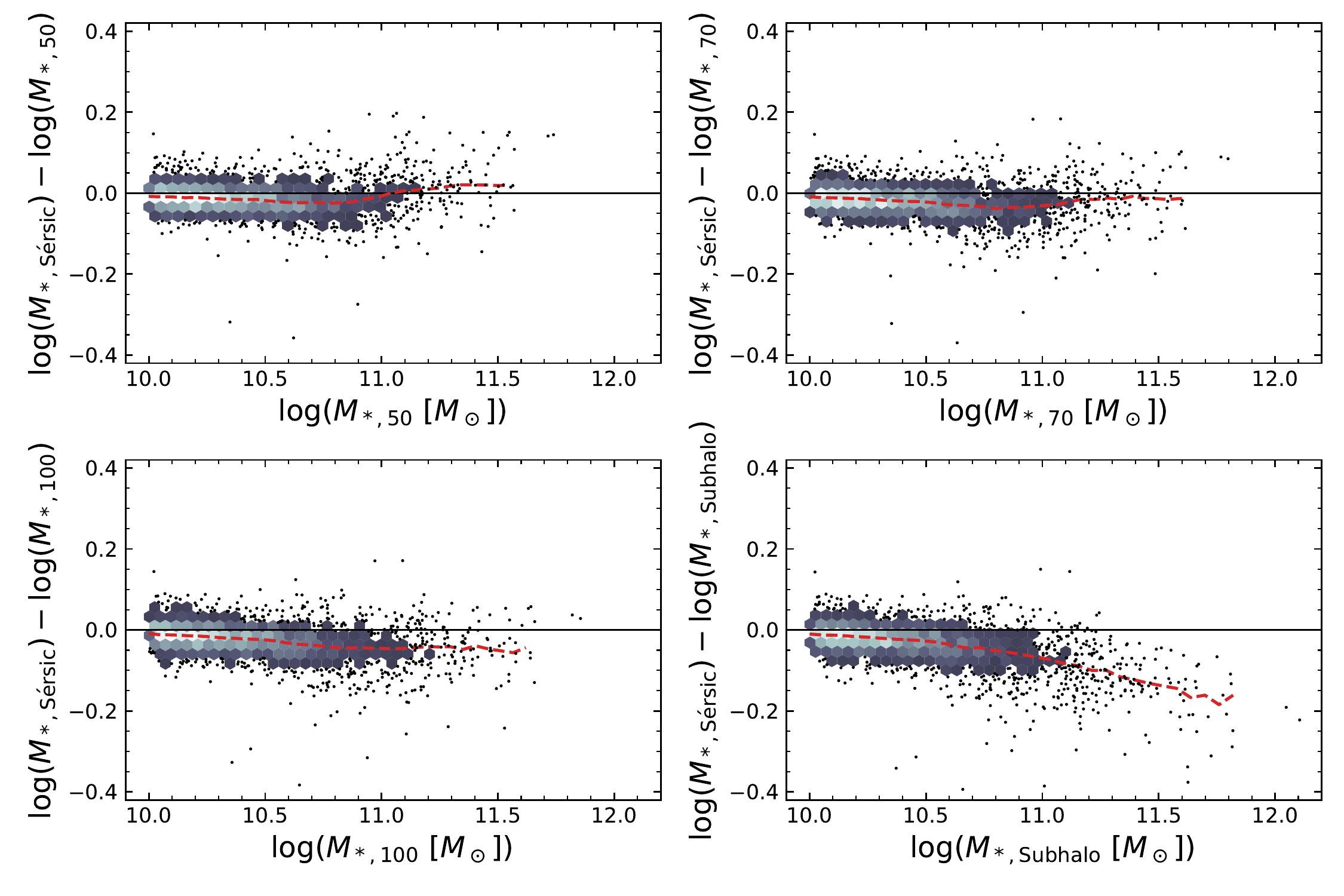}
    \caption{The difference between the total stellar mass of the best-fit S\'ersic model and the mass of the stellar particles enclosed within a specified aperture, as a function of the aperture stellar mass. The top panels, as well as the bottom left panel, show the results for a spherical aperture of radius 50\,kpc, 70\,kpc, and 100\,kpc. Red dashed lines indicate the running median. For comparison, the bottom right panel shows the total mass of all stellar particles belonging to the subhalo. If the aperture is too small (radius of 30\,kpc, Fig.~\ref{fig:flux_recovery}), the S\'ersic model mass deviates strongly from the aperture mass at the at the high mass end. Conversely, the single S\'ersic profile cannot capture all the stellar mass within the subhalo. An aperture of radius of $\sim70$\,kpc appears to give the best agreement between the two different measures of stellar mass, with an approximately constant offset of $-0.02\,$dex across the entire range in stellar mass.}
    \label{fig:aper_comp}
\end{figure*}

In Fig.~\ref{fig:flux_recovery}, we found good agreement between the total stellar mass of the S\'ersic models and the stellar mass measured within a spherical aperture of 30\,kpc, except for the more massive ($M_*\gtrsim 10^{11}\,\rm M_\odot$) galaxies. For the comparison of the population statistics of simulated galaxies with observations, e.g. the mass-size relation or the stellar mass function, it may be of interest to evaluate which aperture best captures the stellar mass at the high mass end. Here, `best' is defined as being as close to what is typically observed, which does not necessarily correspond to the true stellar mass of a galaxy.

To this end, we compare the total stellar mass of the best-fit S\'ersic model (from fits to the mock stellar mass images) with different definitions of stellar mass available in the public EAGLE catalogues. Fig.~\ref{fig:aper_comp} shows this comparison for stellar masses that are calculated as the sum of the stellar particle masses within spherical apertures of increasing radius (50\,kpc, 70\,kpc, 100\,kpc). The bottom right panel compares the S\'ersic model mass with the total stellar mass of the subhalo. Dashed lines show the running median in each panel. 

Contrary to Fig.~\ref{fig:flux_recovery}, where $M_{\rm *,S{\acute e}rsic} > M_{*,30}$ toward high stellar mass, we find that if the aperture is too large (radius of 100\,kpc, or the full subhalo), the S\'ersic model significantly underestimates the total stellar mass at the high mass end. The $M_{*,70}$ mass is in best agreement with $M_{\rm *,S{\acute e}rsic}$ at both lower and high stellar mass (with $M_{*,70}$ being $0.02\,$dex greater on average), suggesting that a spherical aperture of radius $\sim 70\,$kpc will provide a measure that is most consistent with observations.


\bsp	
\label{lastpage}
\end{document}